\documentclass[fleqn,referee]{mnras}
\usepackage{graphicx}
\usepackage[T1]{fontenc}
\usepackage{ae,aecompl}
\def\ph{\phantom}

\title[The s-process enriched star HD 55496]{The s-process enriched star HD
55496: origin from a globular cluster or from the tidal
disruption of a dwarf galaxy? \thanks{Based 
on the observations made with the 2.2m telescope at the European Southern Observatory 
(La Silla, Chile) under the agreement between Observat\'orio Nacional (Brazil) and 
European Southern Observatory (ESO)}}

\author[Pereira et al.]{C.B. Pereira$^{1}$\thanks{E-mail:claudio@on.br}, 
  N.~A.~Drake$^{1,2,3}$\thanks{E-mail:drake@on.br},
  F.~Roig$^{1}$\thanks{E-mail:froig@on.br}
\\ \\
$^{1}$Observat\'orio Nacional/MCTIC, Rua Gen. Jos\'e Cristino, 77, 20921-400, 
Rio de Janeiro, Brazil\\
$^{2}$ Laboratory of Observational Astrophysics, Saint Petersburg State University, 
Universitetski pr. 28, Petrodvoretz 198504, Saint Petersburg,\\ 
Russia\\
$^{3}$ Laborat\'orio Nacional de Astrof\'{\i}sica/MCTIC, Rua Estados Unidos, 154, 37500-000,
Itajub\'a, Brazil\\}

\date{Accepted xxx. Received xxx; in original form xxxx}
\pubyear{2017}

\begin{document}

\label{firstpage}
\pagerange{\pageref{firstpage}--\pageref{lastpage}}

\maketitle

\begin{abstract} 

\par We present a new abundance analysis of HD 55496, previously known
as a metal-poor barium star. We found that HD 55496 has a metallicity
[Fe/H]\,=\,$-1.55$ and is s-process enriched.  We find that HD 55496
presents four chemical peculiarities\,: (i) a Na-O abundance
anti-correlation; (ii) it is aluminum rich; (iii) it is carbon poor
for a s-process enriched star and (iv) the heavy 2$^{nd}$ s-process
peak elements, such as Ba, La, Ce, and Nd, present smaller abundances
than the lighter s-process elements, such as Sr, Y and Zr, which is not
usually observed among the chemically-peculiar binary stars at this
metallicity.  The heavy-element abundance pattern suggests that the
main source of the neutrons is the $^{22}$Ne($\alpha$,n)$^{25}$Mg
reaction.  Taken all these abundance evidence together into
consideration, this strongly suggests that HD 55496 is a ``second
generation of globular cluster star'' formed from gas already strongly
enriched in s-process elements and now is a field halo object.  Our
dynamical analysis, however, indicates that the past encounter
probabilities with the known globular clusters is very small ($\leq
6\%$). This evidence, together with the fact of having a retrograde
motion, points to a halo intruder possibly originated from the tidal
disruption of a dwarf galaxy.

\end{abstract}


\section{Introduction}

\par It is well known that ``second generation of globular cluster
stars'' present an anti-correlation of the abundances among the light
elements, such as a depletion of carbon, oxygen and magnesium
accompanied by an overabundance of other light elements, such as
sodium, aluminum and nitrogen (e.g. Shetrone 1996, Carretta et
al. 2010, just to name a few of the many works dedicated to
investigate the chemical abundances in globular clusters). Such
abundance pattern or the anti-correlations between these abundances
have not been observed in the past among the stars of the halo of low
metallicity.  Thus, any star found in the field halo in the Galaxy
presenting such abundance pattern could be considered as a second
generation candidate star of a globular cluster that has escaped from
the cluster (Fern\'andez-Trincado et al. 2016, Carretta et
al. 2010). Finding these kind of stars within field halo of the Galaxy
is not an easy task and few of them have been found in the last years,
thanks to large spectroscopic surveys (Ram\'{\i}rez et al. 2012;
Schiavon et al. 2017; Fern\'andez-Trincado, 2016, 2017, Martell et
al. 2016), and to photometric plus dynamical analysis searching for
tidal debris from $\omega$ Cen (Majewski et al. 2012).

\par In this work we present a new abundance and dynamical analysis of
the star HD 55496.  This star, first noticed as a barium star by
MacConnel et al. (1972) and later classified as a "metal-poor barium
star" by Luck \& Bond (1991), was observed during the high-resolution
spectroscopic survey dedicated to investigate and to determine the
heavy-abundance abundance pattern of a sample of barium stars. Barium
stars represent among the chemically peculiar stars the largest
sample, and therefore they are useful targets to constrain
nucleosynthesis process in the asymptotic giant branch (AGB) stars (de
Castro et al. 2016, Cseh et al. 2018). Being a barium star, HD 55496
should also be a binary star, but analysis of radial velocity data
collected over 14 years were inconclusive (Jorissen et al. 2015).

\par As we shall see, HD 55496 is an atypical object to be classified
as a barium star. Not only due to the inconclusive results of the
radial velocity variation, but also due to its peculiar chemical
abundance of the elements created by the s-process and the high sodium
and aluminum abundances. This chemical peculiarity, turns HD 55496 to
be an attractive object for an investigation whether it can be a
``second generation of globular cluster star'' which is now a field
halo object.

\par This work is organized as follows: In Section 2 we describe the
observations.  In Section 3 we describe the determination of the
atmospheric parameters and the chemical abundances. In Section 4, we
discuss the results obtained for the luminosity, for the radial
velocity, for the chemical abundances and for the kinematical
analysis. Finally, our conclusions are presented in Section 5.
  
\section{Observations} 

\par The high-resolution spectrum of HD 55496 was obtained with the
FEROS (Fibre-fed Extended Range Optical Spectrograph) echelle
spectrograph (Kaufer et al. 1999) at the 2.2\,m ESO telescope in La
Silla (Chile), during the night of April 3, 2007.  The exposure time
for this observation was 900 sec.  Technical details about the
FEROS spectrograph is reported in Santrich et al. (2013).

\par Since HD 55496 has a high sodium and aluminum abundance it is
important to compare the spectrum of HD 55496 with another star whose
atmospheric parameters are similar to that of HD 55496 and not either
sodium nor aluminum enriched.  In this case, the differences in the
strengths of absorption lines of sodium and aluminum between the two
stars will be only due to a high abundance of sodium and aluminum in
HD 55496.  Figures 1, 2 and 3 show the spectra of HD 55496 in the
spectral regions of the interest of the present work, that is around
the Na\,{\sc i} and Al\,{\sc i} absorption lines in comparison with
the spectrum of CD-27$^\circ$8864. CD-27$^\circ$8864, is a metal-poor
star with similar effective temperature ($T_{\rm eff}$\,=\,4\,650\,K),
surface gravity ($\log g$\,=\,1.6), iron abundance
([Fe/H]\,=\,$-$1.54), and microturbulent velocity 1.6 km s$^{-1}$
analyzed in Pereira et al. (2019).

\begin{figure} %Fig1
\includegraphics[width=\columnwidth]{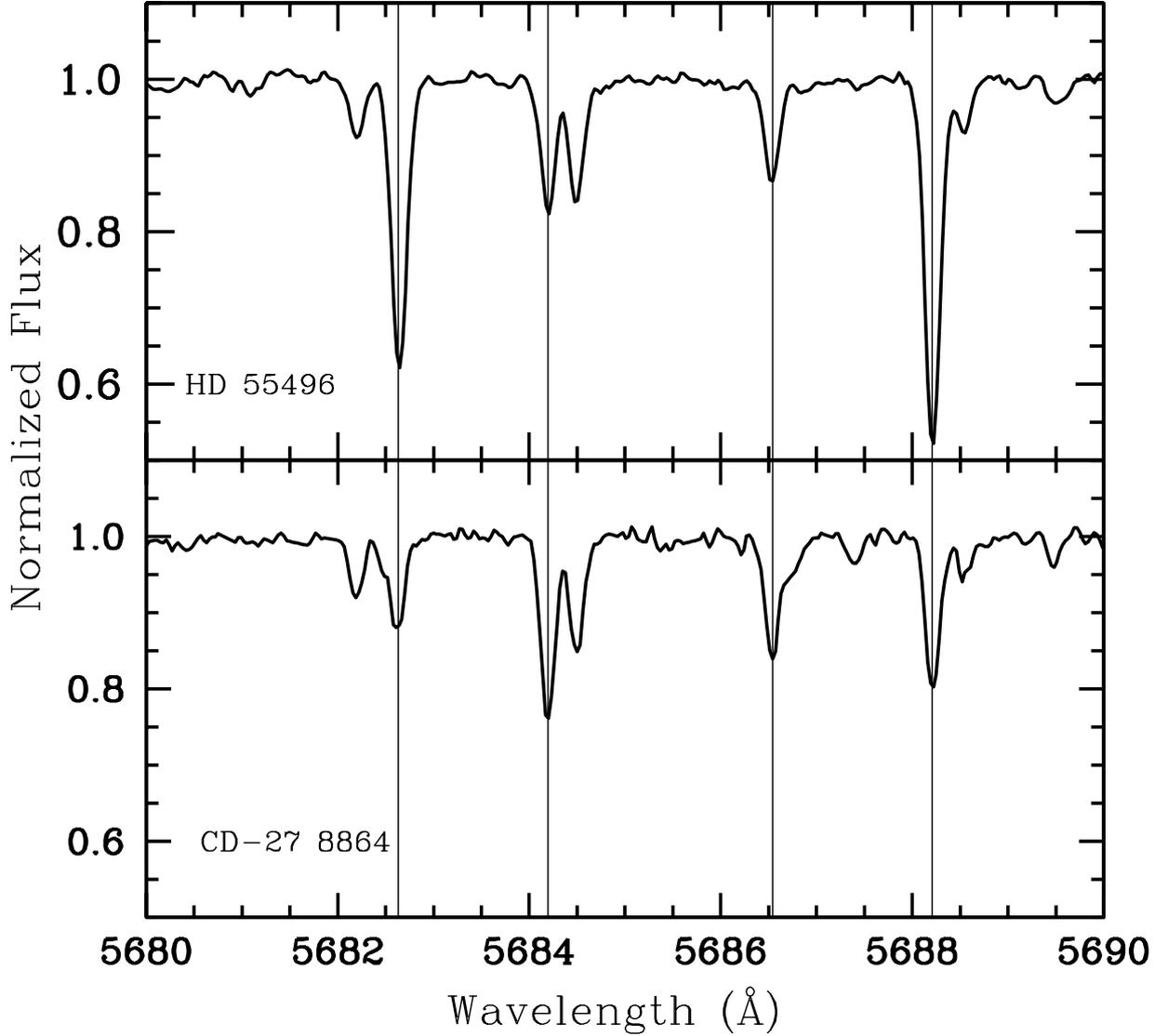}
\caption{Spectra of HD 55496 and CD-27$^\circ$8864 between 5680 and 5690\,\AA\,.
Vertical lines represent the transitions of Na\,{\sc i} 5682.65, Sc\,{\sc ii} 5684.20,
Fe\,{\sc i} 5686.54 and Na\,{\sc i} 5688.22\,\AA\,. We show the spectra of CD-27$^\circ$8864
because it has similar temperature, surface gravity and metallicity of HD 55496.}
\end{figure}

\begin{figure} %Fig2
\includegraphics[width=\columnwidth]{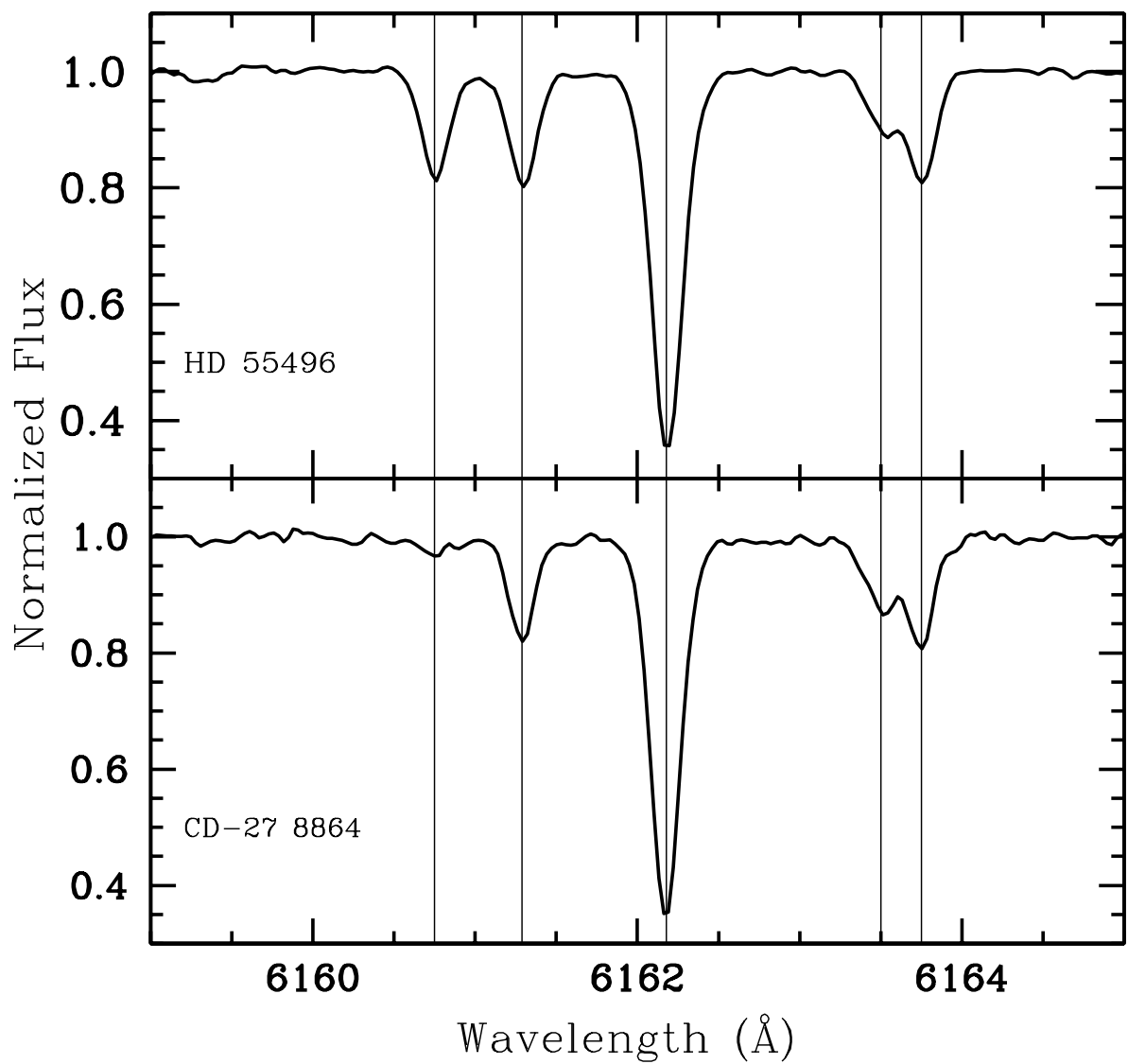}
\caption{Same as Figure 1 but for the spectral region between 6159 and 6165\,\AA\,.
Vertical lines represent the transitions of Na\,{\sc i} 6160.73, Ca\,{\sc i} 6161.27,
6162.18 and Fe\,{\sc i} 6163.55 and Ca\,{\sc i} 6163.75\,\AA.}
\end{figure}

\begin{figure} %Fig3
\includegraphics[width=\columnwidth]{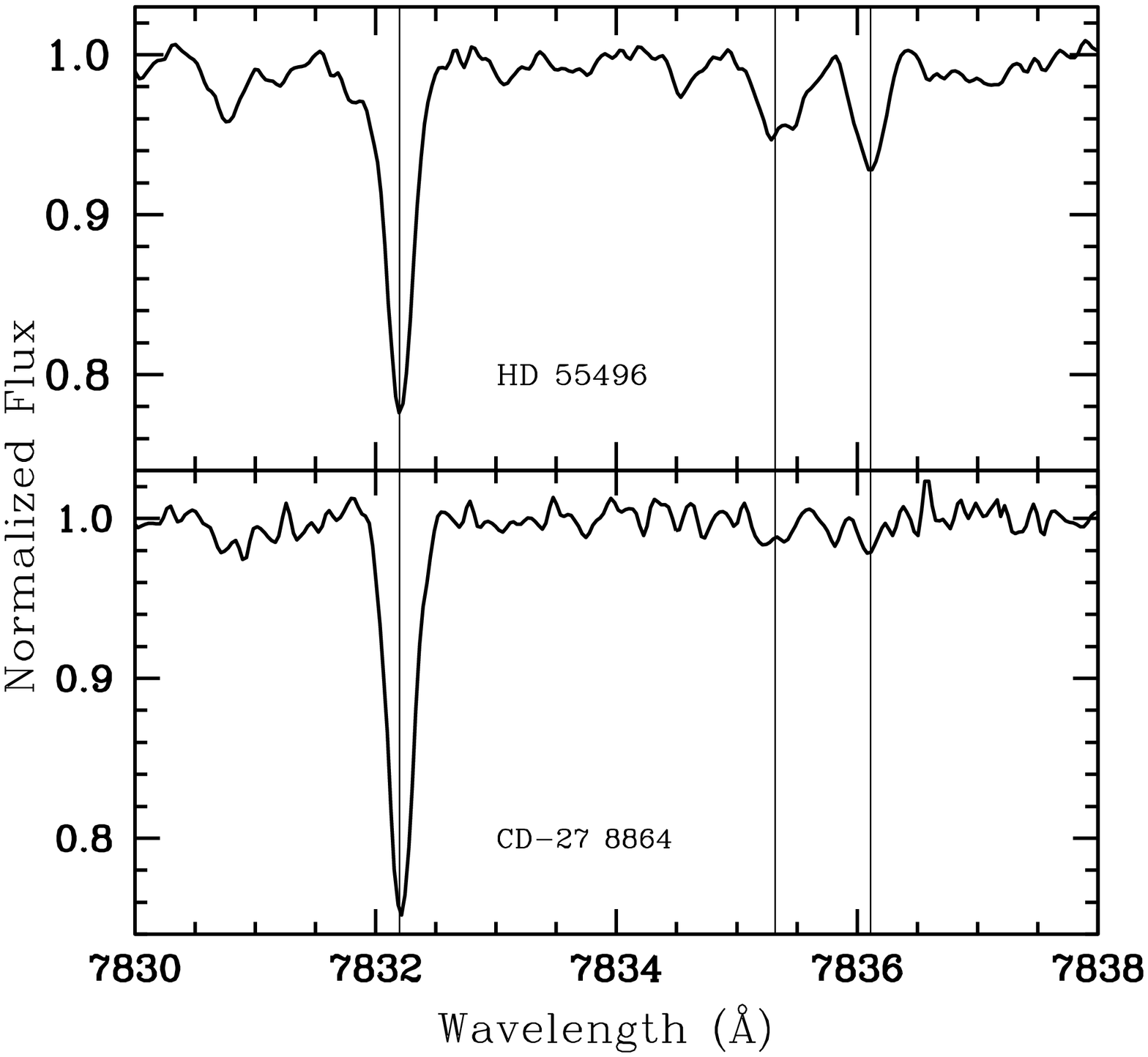}
\caption{Same as Figure 1 but for the spectral region between 7830 and 7838\,\AA\,.
Vertical lines represent the transition of Fe\,{\sc i} at 7838.21\,\AA\, and the  Al\,{\sc i}
transitions at 7835.32 and 7836.13\,\AA.}
\end{figure}

\section{Analysis}

\subsection{Atmospheric Parameters}

\par For the determination of the atmospheric parameters it is first
necessary to measure the equivalent widths of the absorption lines of
Fe\,{\sc i} and Fe\,{\sc ii}. The absorption lines selected for this
task were selected from Lambert et al. (1996) and that was the same
source of lines used in the work of Moriz et al. (2017) in the study
of the CD-50$^\circ$776 recently identified as a new
CEMP-s. \footnote{CEMP stands for Carbon-Enhanced Metal-Poor. The
  suffix $''$s$''$ means that the star is s-process enriched.}  Table
1 shows the values of the equivalent widths of the Fe\,{\sc i} and
Fe\,{\sc ii} absorption lines with their respective excitation
potentials ($\chi$ in eV) and the $\log gf$ values of the transtions.
The equivalent widths were determined using the {\sl splot} task in
IRAF by adjusting Gaussian profiles to the observed
profiles.\addtocounter{table}{1}

\par In addition, the determination of the stellar atmospheric
parameters, such as effective temperature ($T_{\rm eff}$), surface
gravity ($\log g$), microturbulence ($\xi$), and metallicity ([Fe/H])
(we use the notation [X/H]=$\log(N_{\rm X}/N_{\rm H})_{\star} -
\log(N_{\rm X}/N_{\rm H})_{\odot}$), were also done in the same way as
in Roriz et al. (2017).  In brief, it consists in using the local
thermodynamic equilibrium (hereafter LTE) model atmospheres of Kurucz
(1993) and the spectral analysis code {\sc moog}, version 2013 (Sneden
1973). Table 2 shows the final adopted atmospheric parameters with
their respective errors.  Previous atmospheric parameters
determination for HD 55496 were obtained by Karinkuzhi \& Goswami
(2015) and earlier by Luck \& Bond (1991). Our results are similar to
those of Karinkuzhi \& Goswami (2015).

\par The errors in temperature and microturbulent velocity were
estimated considering, respectively, the uncertainty in the value of
the slope of the relationships between the abundance of iron, given by
the Fe\,{\sc i} lines, and the excitation potential and the abundance
of iron and the reduced equivalent width.  For gravity, the
uncertainty was estimated considering that the difference between the
Fe\,{\sc i} and Fe\,{\sc ii} abundances differed by 1$\sigma$ of the
standard deviation of the mean value of Fe\,{\sc i} abundances. We
found for the temperature, surface gravity and for the microturbulent
velocity uncertainties of $\sigma(T_{\rm eff}) = \pm 70$ $\sigma(\log
g)\!=\!\pm 0.2$ and $\sigma(\xi) = \pm 0.3$ km\,s$^{-1}$,
respectively.

\begin{table} %T2
\caption{Atmospheric parameters for HD 55496.}
\begin{tabular}{lcc}\hline
Parameter & Value &  Ref. \\\hline
$T_{\rm eff}$ (K)        & 4\,700$\pm$70    & 1  \\
                       & 4\,850           & 2  \\
                       & 4\,800           & 3  \\
                       & 4\,750           & 4  \\\hline
                       
$\log g$ (dex)         & 1.9$\pm$0.2      & 1  \\
                       & 2.05             & 2  \\
                       & 2.8              & 3  \\
                       & 2.5              & 4  \\\hline

[Fe/H] (dex)           & $-$1.55$\pm$0.11   & 1 \\
                       & $-$1.49            & 2 \\
                       & $-$1.55            & 3 \\\hline

$\xi$ (km\,s$^{-1}$)    & 1.5$\pm$0.3       & 1 \\
                       & 1.5               & 2 \\
                       & 2.7               & 3 \\
                       & 1.5               & 4 \\\hline

\end{tabular}
\par References for Table 2.
\par 1: This work;
\par 2: Karinkuzhi \& Goswami (2015);
\par 3: Luck \& Bond (1991);
\par 4: Sneden (1983);
\end{table}

\subsection{Abundance analysis}                                        

\par The abundances of chemical elements Na, Mg, Al, Si, Ca, Ti, Cr,
Ni, Sr, Y, Zr, Mo, Ce, Nd and Sm were determined by measuring the
equivalent widths of their absorption lines considering
local-thermodynamic-equilibrium (LTE) model-atmosphere.  We used the
line-synthesis code {\sc moog}, (Sneden, 1973) for the calculations
and Table 3 shows the atomic lines used to obtain the abundances of
the elements and Table 4 provides the results, the number of lines
employed for each species, $n$, and the standard deviations. Our
abundances were determined differentially with respect to the solar
abundances of Grevesse \& Sauval (1998).\addtocounter{table}{1}

\par The abundances of carbon, nitrogen, oxygen and lithium were
determined by means of the spectral synthesis technique.  For carbon,
the abundance was determined using the CH lines of the $A^2\Delta -
X^2Pi$ system at $\sim 4365$\,\AA\.  For the determination of the
abundance of nitrogen and the $^{12}$C/$^{13}$C isotopic ratio, we
used the $^{12}$CN lines of the (2,\,0) band of the CN red system
$A^2\Pi - X^2\Sigma$ in the 7994 -- 8020~\AA\ wavelength range. For
the abundance of oxygen, we used the forbidden line at 6300.3\,\AA\,
where we adopted $\log gf$\,=\,$-$9.72 from Allende Prieto et
al. (2001).  The lithium abundance was derived using the Li\,{\sc i}
$\lambda$ 6708\,\AA\, resonance doublet.

\par The oscillator strengths and wavelengths of the CH and CN
molecular lines and the Li\,{\sc i} doublet were taken from Drake \&
Pereira (2008). For the other elements which were abundances were
determined using spectral synthesis, such as barium, lanthanum, and
lead, the line lists were constructed using the databases of
Kurucz\footnote{http://kurucz.harvard.edu/} and
VALD.\footnote{http://vald.astro.uu.se/} Figures 4, 5 and 6 show the
observed and synthetic spectra for the spectral regions where the
abundances of carbon, nitrogen and lead were obtained.

\par The abundances of barium, lanthanum and lead were also determined
by means of the spectral synthesis technique. The determination of
barium abundance was obtained using the Ba\,{\sc ii} lines at $\lambda
4554.0$, $\lambda 4934.1$ and $\lambda 5853.7\,\AA$.  Hyperfine and
isotope splitting were taken from McWilliam (1998) and the lead
abundance was derived from the Pb\,{\sc i} line at $\lambda
4057.81$\,\AA\, where the isotopic shifts and hyperfine splitting were
taken from van Eck et al. (2003).  For lanthanum, the abundance was
determined using the La\,{\sc ii} lines at $\lambda 4086.7$, $\lambda
4662.5$, $\lambda 5114.5$ and $\lambda 5303.5$ where the isotopic
shifts and hyperfine splitting were taken from Roederer \& Thompson
(2015) and Karinkuzhi et al. (2018).

\begin{figure} %Fig4
\includegraphics[width=\columnwidth]{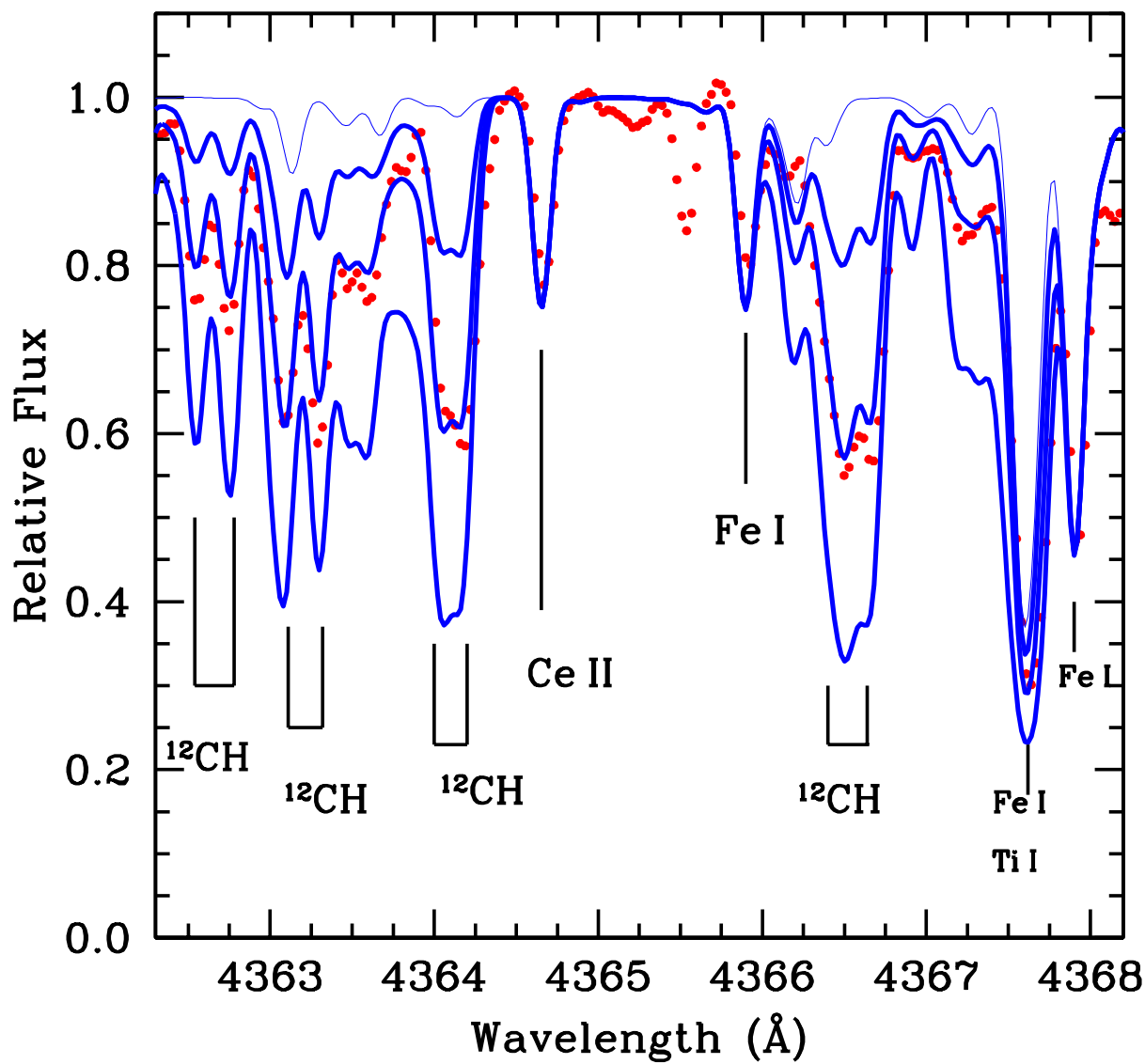}
\caption{The observed spectrum of HD 55496 (red dots) and synthetic spectra
  in the spectral region between 4362.3\,\AA\, and 4368.2\,\AA\,
where we can see the CH molecular lines. The lighter blue line shows
the synthetic spectrum calculated without the contribution of the CH
molecular lines.  Blue solid lines, from top to bottom, show three
synthetic spectra calculated respectively for the carbon abundances equal
to log$\epsilon$(C)\,=\,6.12, 6.62 (adopted in this work) and 7.12. Other absorptions
lines are also showed.}
\end{figure}

\begin{figure} %Fig5
\includegraphics[width=\columnwidth]{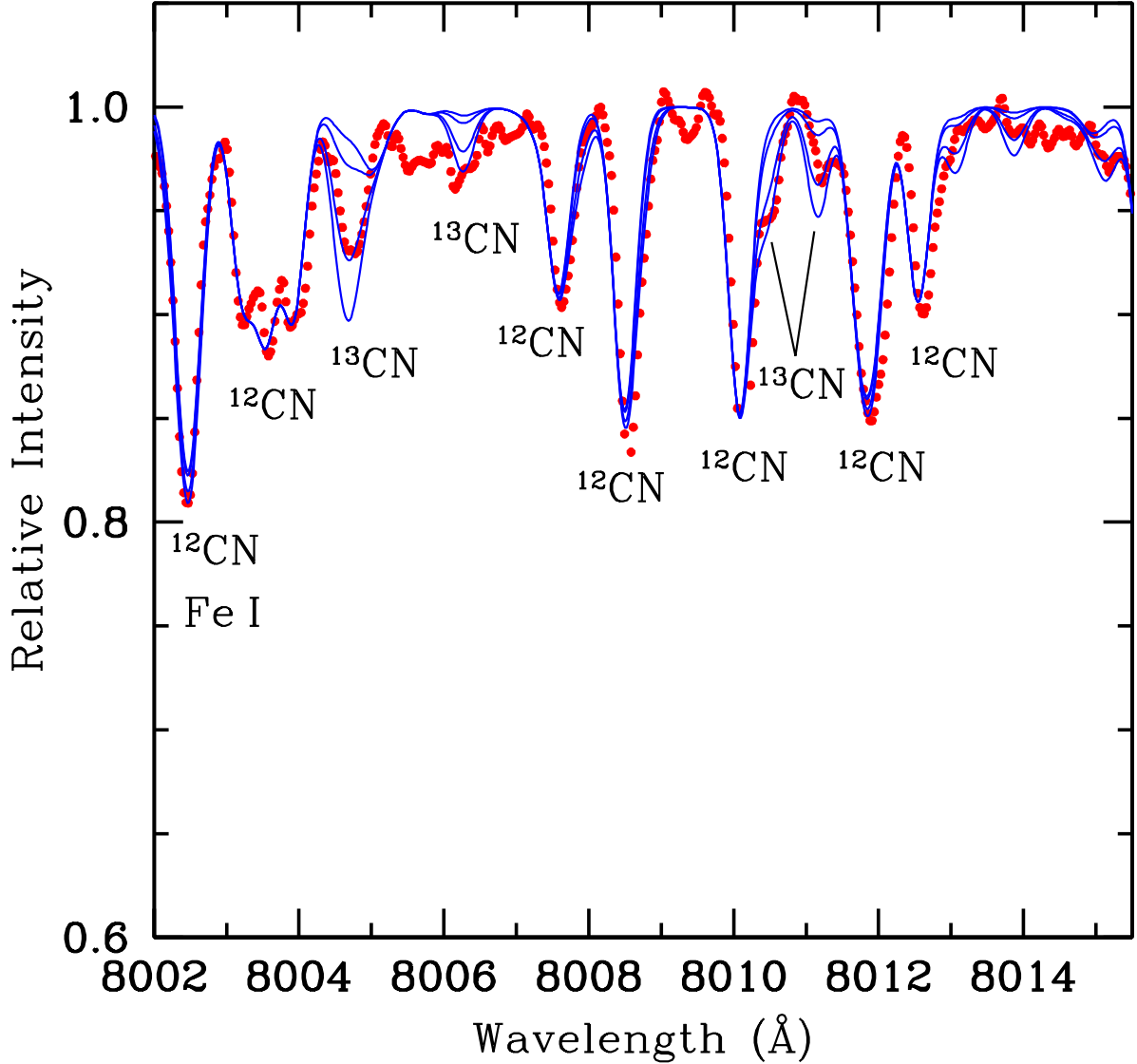}
\caption{Observed (dotted red points) and synthetic (solid blue lines)
spectra between $\lambda$ 8002 and $\lambda$ 8014.5\,\AA.\, 
From top to bottom we show the syntheses for four values of the $^{12}$C/$^{13}$C
isotopic ratio (36.0, 18.0, 6.0, and 4.0) and for 
$\log\varepsilon$(C)\,=\,6.62, $\log\varepsilon$(N)\,=\,8.17
(adopted) and $\log\varepsilon$(O)\,=\,7.43. Other $^{12}$CN absorptions are also indicated.}
\end{figure}

\begin{figure} %Fig6
\includegraphics[width=\columnwidth]{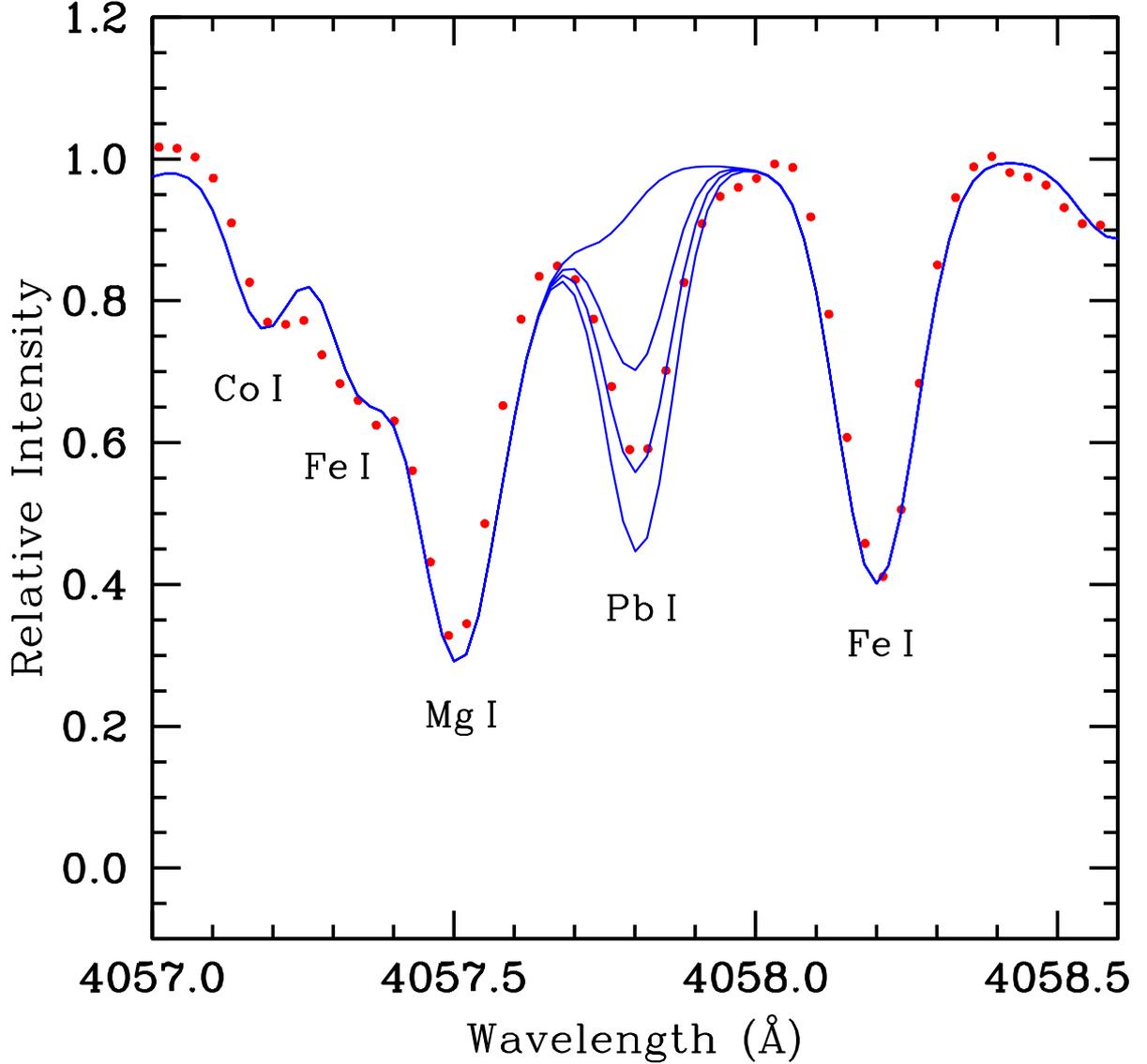}
\caption{The observed spectrum of HD 55496 (red dots)
in the spectral region between 4057.0\,\AA\, and
4058.6\,\AA\, where we can see the transition of Pb\,{\sc i} at 4057.8\AA\,. 
Synthetic spectra (solid blue lines) are also shown calculated
respectively for the lead abundances equal
to log$\epsilon$(Pb)\,=\,1.25, 1.75 (adopted in this work) and 2.25.
The lighter blue line shows the synthetic spectrum calculated
without the contribution of the lead. Other absorptions
lines are also showed.}
\end{figure}

\begin{table} %Tab 4
\caption{Elemental abundances derived for HD 55496. The second
column provides the solar abundances. The third column gives the
information whether abundances were determined using spectrum
synthesis technique (syn) or based on the equivalent width
measurements.  In this latter case we provide the number of lines
(n) used for the abundance determination.  The fourth and fifth
columns give, respectively, the abundance in the scale
$\log\varepsilon$({\rm H}) = 12.0 with the respective abundance
dispersion among the lines of the elements with more than three
available lines. The sixth and the seventh columns give the
abundances in the notations [X/H] and [X/Fe]. For the
elements Sr, Mo and Pb the abundance dispersion was calculated
considering three different positions of the continuum.}
\begin{tabular}{lcccccc}\hline\hline
Species & $\log\varepsilon$$^{\odot}$ & n(\#) & $\log\varepsilon$ & $\sigma_{\rm obs}$ & [X/H] & [X/Fe]\\\hline 
Li           & 3.31 & syn & 0.00 & 0.15 &    ---  &    ---  \\
C\,(CH)      & 8.52 & syn & 6.62 & 0.05 & $-$1.90 & $-$0.35 \\
N\,(CN)      & 7.92 & syn & 8.17 & 0.05 & $+$0.25 & $+$1.80 \\
O            & 8.83 & syn & 7.43 & 0.10 & $-$1.40 & $+$0.15 \\
Na\,{\sc i}  & 6.33 &  8  & 5.59 & 0.13 & $-$0.74 & $+$0.81 \\
Mg\,{\sc i}  & 7.58 &  5  & 6.73 & 0.14 & $-$0.85 & $+$0.70 \\
Al\,{\sc i}  & 6.47 &  6  & 5.66 & 0.12 & $-$0.81 & $+$0.74 \\
Si\,{\sc i}  & 7.55 &  6  & 6.66 & 0.09 & $-$0.89 & $+$0.66 \\
Ca\,{\sc i}  & 6.36 & 16  & 5.39 & 0.12 & $-$0.97 & $+$0.58 \\
Ti\,{\sc i}  & 5.02 & 24  & 3.72 & 0.07 & $-$1.30 & $+$0.25 \\
Fe\,{\sc i}  & 7.50 & 70  & 5.95 & 0.11 & $-$1.55 &  --- \\
Fe\,{\sc ii} & 7.50 & 10  & 5.92 & 0.09 & $-$1.58 &  --- \\
Cr\,{\sc i}  & 5.67 & 10  & 4.06 & 0.11 & $-$1.61 & $-$0.06 \\
Ni\,{\sc i}  & 6.25 & 35  & 4.70 & 0.09 & $-$1.55 &    0.00 \\
Sr\,{\sc i}  & 2.97 &  1  & 2.04 & 0.07 & $-$0.93 & $+$0.62 \\
Y\,{\sc ii}  & 2.24 &  7  & 1.53 & 0.13 & $-$0.71 & $+$0.84 \\
Zr\,{\sc i}  & 2.60 &  6  & 2.20 & 0.06 & $-$0.40 & $+$1.15 \\
Zr\,{\sc ii} & 2.60 &  4  & 2.28 & 0.06 & $-$0.32 & $+$1.23 \\
Mo\,{\sc i}  & 1.92 &  1  & 1.47 & 0.10 & $-$0.45 & $+$1.10 \\
Ba\,{\sc ii} & 2.13 & syn & 1.20 & 0.06 & $-$0.93 & $+$0.62 \\
La\,{\sc ii} & 1.17 & syn & $-$0.07 & 0.08 & $-$1.24 & $+$0.31 \\
Ce\,{\sc ii} & 1.58 & 10  & 0.54 & 0.10 & $-$1.04 & $+$0.51 \\
Nd\,{\sc ii} & 1.50 & 21  & 0.34 & 0.10 & $-$1.16 & $+$0.39 \\
Sm\,{\sc ii} & 1.01 &  6  & $-$0.35 & 0.06 & $-$1.36 & $+$0.19 \\
Pb\,{\sc i}  & 1.95 & syn & 1.75 & 0.05  & $-$0.20 & $+$1.35 \\\hline
\end{tabular}
\end{table}

\subsection{Abundance uncertainties}    

\par Table 5 shows the uncertainties in the abundances of the chemical
elements due to errors in temperature, surface gravity,
microturbulence and metallicity determined previously in Section 3.1.
Abundance errors were estimated by changing these parameters
individually according to the uncertainties shown in Table 2.  The
uncertainties in the measurements of equivalent widths were estimated
using the expression given by Cayrel (1988). For the Feros spectrum
having a nominal resolution of 48\,000 and a S/N around 100, the
estimated error is of the order of 3\,m\AA,. These error estimates
were then taken into account in the measurements of the equivalent
widths of the absorption lines of the chemical elements, provinding
other values for the elemental abundances.  The eighth column of Table
5 shows the final uncertainty due to uncertainties in atmospheric
parameters (temperature, surface gravity, metallicity and
microturbulence velocity), equivalent widths and the dispersion of the
abundances.  The total uncertainty was calculated as the root
quadratic sum of each of these parameters considering that each
individual uncertainty is independent.

\par For the elements analyzed through spectral synthesis technique we
used the same methodology that was used to calculate the uncertainties
of the abundances obtained based on measurements of the equivalent
widths.  We vary the temperature, surface gravity, metallicity and
microturbulence velocity and then compute independently the variation
of the abundances obtained by each value.  In addition, for carbon,
nitrogen and oxygen, we consider that their respective abundances are
interdependent, that is, the uncertainty in the determination of
abundance of oxygen affects the abundance of carbon and
vice-versa. The uncertainties in carbon abundance also affect the
abundance of nitrogen since the CN molecule was used for the
determination of nitrogen abundance.  For the light elements, carbon,
nitrogen and oxygen, the variations of the abundance due to changes in
effective temperature, surface gravity and micro-turbulent velocities
are given in Table 6.  Abundances based on molecular lines are little
affected by microturbulence velocity variation because they are weak
lines. As for surface gravity, the molecules can be affected since the
molecular equilibrium is sensitive to the gas pressure.  However,
considering variations of only 0.2 dex for the surface gravity, the
error introduced in abundance is not significant.  In fact similar
uncertainties were found by Tautvai$\check{\rm s}$ien\.e et
al. (2000).

\par Table~5 shows the well known behavior that the neutral elements
are more sensitive to the temperature variations, while singly-ionized
elements are more sensitive to the surface gravity variations (Gray
1992).  For the elements whose abundance is based on stronger lines,
such as strontium, yttrium and calcium the error introduced by the
microturbulence is important.  This is because the equivalent widths
of the moderately strong lines which lie in the saturated part of the
curve of growth are larger than those synthetic lines calculated
considering only thermal and damping broadening. Once microturbulent
velocity is introduced in the line computation, it delays the
saturation of an absorption line in the curve of growth and also
becomes the major source of uncertainty in the abundance analysis
based in these strong lines.  For stronger absorption lines with
saturated cores, the wings grow increasing the line strength and
become responsible for the most of their equivalent widths. Since the
equivalent width is also proportional to the Doppler broadening which
is related to microturbulent velocity, one should expect that
microturbulence also affects the line strength.

\begin{table*}%T5
\caption{Abundance uncertainties for HD 55496. Second column gives
$\sigma_{\rm log\varepsilon}$\,=\,$\sigma_{\rm obs}$/$\sqrt{n}$ where $n$
is the number of absorption lines used for the abundance determination.
See fifth column of Table 4 for the values of $\sigma_{\rm obs}$.
Columns 3 to 7 give the 
variation of the abundances caused by the variation in $T_{\rm eff}$, $\log g$, $\xi$, 
[Fe/H], and equivalent width measurements ($W_\lambda$), respectively. 
The 8th column gives all the uncertainties 
quadratically combined from the 3rd to 7th columns. The last column gives the 
abundance dispersion observed among the lines for those elements with three
of more lines available as already shown in Table 4.}
\begin{tabular}{lcccccccc}\hline
Species & $\sigma_{\rm log\varepsilon}$ & $\Delta T_{\rm eff}$ & $\Delta\log g$ & $\Delta\xi$ & $\Delta$[Fe/H] & 
$\Delta W_{\lambda}$ & $\left( \sum \sigma^2 \right)^{1/2}$ & $\sigma_{\rm obs}$\\
& $_{\rule{0pt}{8pt}}$ & $\pm$70\,K & $\pm$0.2 & $\pm$0.3 km\,s$^{-1}$ & $\pm$0.1 & $\pm$3 m\AA &  \\\hline
Li             & 0.15 & $+$0.10/$-$0.10 & $+$0.05/$-$0.05 &    0.00/0.00    &    0.00/0.00    & ---             & 0.19/0.19 & 0.15 \\
Fe\,{\sc i}    & 0.01 & $+$0.09/$-$0.09 & $-$0.01/$+$0.01 & $-$0.11/$+$0.13 & $-$0.01/$+$0.01 & $+$0.07/$-$0.07 & 0.16/0.17 & 0.11 \\
Fe\,{\sc ii}   & 0.03 & $-$0.02/$+$0.04 & $+$0.09/$-$0.07 & $-$0.03/$+$0.05 & $+$0.02/$-$0.01 & $+$0.09/$-$0.09 & 0.10/0.12 & 0.09 \\
Na\,{\sc i}    & 0.05 & $+$0.06/$-$0.06 & $-$0.02/$+$0.02 & $-$0.05/$+$0.05 & $-$0.01/$+$0.01 & $+$0.05/$-$0.06 & 0.09/0.11 & 0.13 \\
Mg\,{\sc i}    & 0.06 & $+$0.07/$-$0.06 & $-$0.04/$+$0.04 & $-$0.07/$+$0.07 &    0.00/$+$0.01 & $+$0.05/$-$0.04 & 0.11/0.12 & 0.14 \\
Al\,{\sc i}    & 0.05 & $+$0.04/$-$0.05 & $-$0.01/$+$0.01 & $-$0.02/$+$0.02 & $-$0.01/$+$0.01 & $+$0.06/$-$0.07 & 0.09/0.10 & 0.12 \\
Si\,{\sc i}    & 0.04 & $+$0.02/$-$0.01 & $+$0.02/$-$0.01 & $-$0.01/$+$0.02 &    0.00/0.00    & $+$0.09/$-$0.10 & 0.10/0.11 & 0.09 \\
Ca\,{\sc i}    & 0.03 & $+$0.07/$-$0.07 & $-$0.03/$+$0.02 & $-$0.12/$+$0.12 & $-$0.02/$+$0.01 & $+$0.06/$-$0.07 & 0.16/0.16 & 0.12 \\
Ti\,{\sc i}    & 0.01 & $+$0.12/$-$0.13 & $-$0.02/$+$0.02 & $-$0.06/$+$0.07 & $-$0.01/$+$0.02 & $+$0.07/$-$0.07 & 0.15/0.17 & 0.07 \\
Cr\,{\sc i}    & 0.03 & $+$0.10/$-$0.10 & $-$0.02/$+$0.01 & $-$0.08/$+$0.10 & $-$0.01/$+$0.02 & $+$0.08/$-$0.09 & 0.13/0.17 & 0.11 \\
Ni\,{\sc i}    & 0.01 & $+$0.08/$-$0.09 & $-$0.02/0.00    & $-$0.05/$+$0.06 & $-$0.01/0.00    & $+$0.07/$-$0.08 & 0.12/0.13 & 0.09 \\
Sr\,{\sc ii}   & 0.07 & $+$0.12/$-$0.13 & $-$0.03/$+$0.03 & $-$0.17/$+$0.24 & $-$0.02/$+$0.02 & $+$0.08/$-$0.08 & 0.23/0.30 & 0.07 \\
Y\,{\sc ii}    & 0.05 & $+$0.02/$-$0.01 & $+$0.06/$-$0.06 & $-$0.17/$+$0.22 & $+$0.02/$-$0.01 & $+$0.08/$-$0.07 & 0.21/0.24 & 0.13 \\
Zr\,{\sc i}    & 0.02 & $+$0.13/$-$0.13 & $-$0.01/$+$0.02 & $-$0.01/$+$0.02 &    0.00/$+$0.02 & $+$0.10/$-$0.11 & 0.16/0.17 & 0.06 \\
Zr\,{\sc ii}   & 0.03 & $+$0.01/0.00    & $+$0.07/$-$0.06 & $-$0.13/$+$0.19 & $+$0.02/$-$0.01 & $+$0.08/$-$0.07 & 0.17/0.21 & 0.06 \\
Mo\,{\sc i}    & 0.10 & $+$0.11/$-$0.12 & $-$0.02/$+$0.01 & $-$0.01/$+$0.01 & $-$0.01/$+$0.01 & $+$0.13/$-$0.18 & 0.20/0.24 & 0.10 \\
Ba\,{\sc ii}   & 0.03 &    0.01/0.00    & $+$0.10/$-$0.10 & $-$0.20/$+$0.30 &    0.00/0.00    &  ---            & 0.23/0.31 & 0.06\\
La\,{\sc ii}   & 0.04 & $-$0.05/$+$0.05 & $+$0.10/$-$0.10 &    0.00/0.00    &    0.00/0.00    &  ---            & 0.12/0.12 & 0.08 \\
Ce\,{\sc ii}   & 0.03 & $+$0.02/$-$0.02 & $+$0.08/$-$0.07 & $-$0.05/$+$0.09 & $+$0.02/$-$0.02 & $+$0.08/$-$0.09 & 0.13/0.15 & 0.10 \\
Nd\,{\sc ii}   & 0.02 & $+$0.02/$-$0.03 & $+$0.07/$-$0.08 & $-$0.04/$+$0.05 & $+$0.02/$-$0.03 & $+$0.08/$-$0.09 & 0.12/0.14 & 0.10 \\
Sm\,{\sc ii}   & 0.02 & $+$0.03/$-$0.02 & $+$0.08/$-$0.07 & $-$0.02/$+$0.04 & $+$0.03/$-$0.02 & $+$0.09/$-$0.09 & 0.13/0.13 & 0.06 \\
Pb\,{\sc i}    & 0.05 & $+$0.10/$-$0.20 &    0.00/0.00    & $-$0.10/$+$0.10 &    0.00/0.00    &  ---            & 0.11/0.22 & 0.05  \\\hline

\end{tabular}
\end{table*}

\begin{table*} %T6
\caption{Abundance errors for the elements C, N, O of HD 55496.} 
\begin{tabular}{ccccccccc}\hline 
  Species  & $\sigma_{\rm log\varepsilon}$ & $\Delta T_{\rm eff}$ & $\Delta\log g$ & $\Delta\xi$ & $\Delta\log {\rm (C)}$
   & $\Delta\log{\rm (N)}$ & $\Delta\log{\rm (O)}$ & $\sigma_{\rm tot}$ \\ 
   &         & $\pm$70~K        &   $\pm$0.2          & $\pm$0.3~km\,s$^{-1}$ & $+$0.20   & $+$0.20 & $+$0.20  &  \\\hline 
C  & 0.05    & $+$0.10/$-$0.10  &      0.00/$-$0.05   &    0.00/0.00         & ---       & 0.00    & $+$0.15  & $+$0.19/$+$0.19\\
N  & 0.02    & $+$0.10/$-$0.10  &   $+$0.05/$-$0.05   &    0.00/0.00         & $-$0.10   & ---     & $+$0.20  & $+$0.29/$+$0.29\\
O  & 0.10(?) & $+$0.05/$-$0.05  &   $+$0.10/$-$0.10   &    0.00/$-$0.05      &    0.00   & 0.00    &  ---     & $+$0.15/$+$0.16\\
\hline                     
\end{tabular}
\end{table*}

\section{Results and Discussion}

\subsection{Luminosity}

\par The luminosity of HD 55496 can be estimated based on the recently
Gaia DR2 distance determination which is 494$\pm$11 pc (Bailer-Jones
2018).  Considering the relationship by Alonso et al. (1999) between
bolometric correction (BC) and effective temperature, BC results to be
$-$0.38.  Assuming $V=8.4$ and $M_{{\rm bol} \odot} = +4.74$ (Bessel,
1998), we obtain log\,($L_{\star}/L_{\odot})$\,=\,2.14\,-\,2.38 or
$L_{\star}$\,=\, 140\,-\,240\,L$_{\odot}$. This range in luminosity is
due to uncertain interstellar absorption because the relatively low
galactic latitude ($b$\,=\,$-$5.96$^\circ$) of HD 55496.  In fact,
using the calibrations between $A_{\rm V}$, galactic coordinates and
distances given by Chen et al. (1998), we obtain $A_{\rm V}$\,=\,0.16
or 0.61.

\par The luminosity above obtained is not enough for a star develop
helium-burning through the thermal pulses during AGB phase and then
becoming a star rich in elements formed by slow neutron
capture. Theoretical calculations show that for a star to develop the
first thermal pulse the required luminosity should be around
1\,800\,$L_{\odot}$ (Lattanzio, 1986) or 1\,400\,$L_{\odot}$
(Vassiliadis \& Wood, 1993).

\par We can also estimate the distance through the derived effective
temperature and surface gravity.  The relation between the distance of
a star to the Sun, $r$, and its temperature, gravity, mass, $V$
magnitude, bolometric correction ($BC$), and interstellar absorption
($A_{V}$), is given by:

\begin{eqnarray}
\log r\: ({\rm kpc}) & = & \frac{1}{2}\left(\log \frac{M_{\star}}{M_{\odot}}
+ 0.4\times\left(V-A_{\rm V}+BC\right) \right. \nonumber \\
& & \left. {{\,}\atop{\,}} + 4\times\log T_{\rm eff} - \log g - 16.5\right).
\end{eqnarray}

\par Using the values of the temperature and surface gravity given in
Table 2 and the values of the $V$-magnitude and the bolometric
correction given above and assuming a mass of $M\,=\,0.8\,M_{\odot}$
for HD 55496 (the low metallicity suggests an old and hence
  lower mass nature) equation (1) can be written as
 
\begin{equation}
\log r\: ({\rm kpc})\,=\,\frac{1}{2}\left(-0.4\times A_{\rm V}\,-\,0.6)\right.
\end{equation}

\par Considering the three possible values for the interstellar
absorption above mentioned 0.16 and 0.61 the obtained distance is,
respectively, 465 and 378 parsecs. The first value, obtained for
A$_{\rm V}$\,=\,0.16 is in agreement with the recent distance
determination of the Gaia DR2 release.

\subsection{Radial Velocity}

\par Table 7 shows all known measurements of radial velocity of HD
55496 available in the literature and determined in this work.  We
determined the radial velocity measuring the Doppler shifts of
selected absorption lines. It is not clear whether the radial velocity
of HD 55496 presents variations due to orbital motion. A similar
conclusion has already been raised by Jorissen et al. (2005).  In
Figure 7 we show the radial velocity measurements obtained by Jorissen
et al. (2005) where we also include our value of 315.63$\pm$0.23
km\,s$^{-1}$.  It seems that there is no significant variation of the
radial velocity.  Systematic radial velocity monitoring is necessary
to confirm the possible binary nature of HD 55496. If HD 55496 would
indeed be a binary star, the observed s-process overabundance could be
due to a mass transfer happened in the past, since HD 55496 is not
luminous enough to be an AGB star.
  
\begin{table} %T7
\caption{Radial velocity (RV) for HD 55496.}
\begin{tabular}{lcc}\hline
RV (km\,s$^{-1}$) & Ref. \\\hline
 315.63$\pm$0.23  & 1 \\
 316.21$\pm$0.16  & 2 \\
 315.28$\pm$0.80  & 3 \\
 322              & 4 \\
 315.30$\rightarrow$317.72 & 5 \\
 317$\pm$10       & 6 \\
 317              & 7 \\\hline

\end{tabular}
\par References for Table 7.
\par 1: This work;
\par 2  GAIA DR2, Bailer-Jones (2018);
\par 3: Karinkuzhi \& Goswami (2015);
\par 4: Gontcharov (2006);
\par 5: Jorissen et al. (2005);
\par 6: Beers \& Sommer-Larsen (1995);
\par 7: Luck \& Bond (1991);
\end{table}

\begin{figure} %Fig7
\includegraphics[width=\columnwidth]{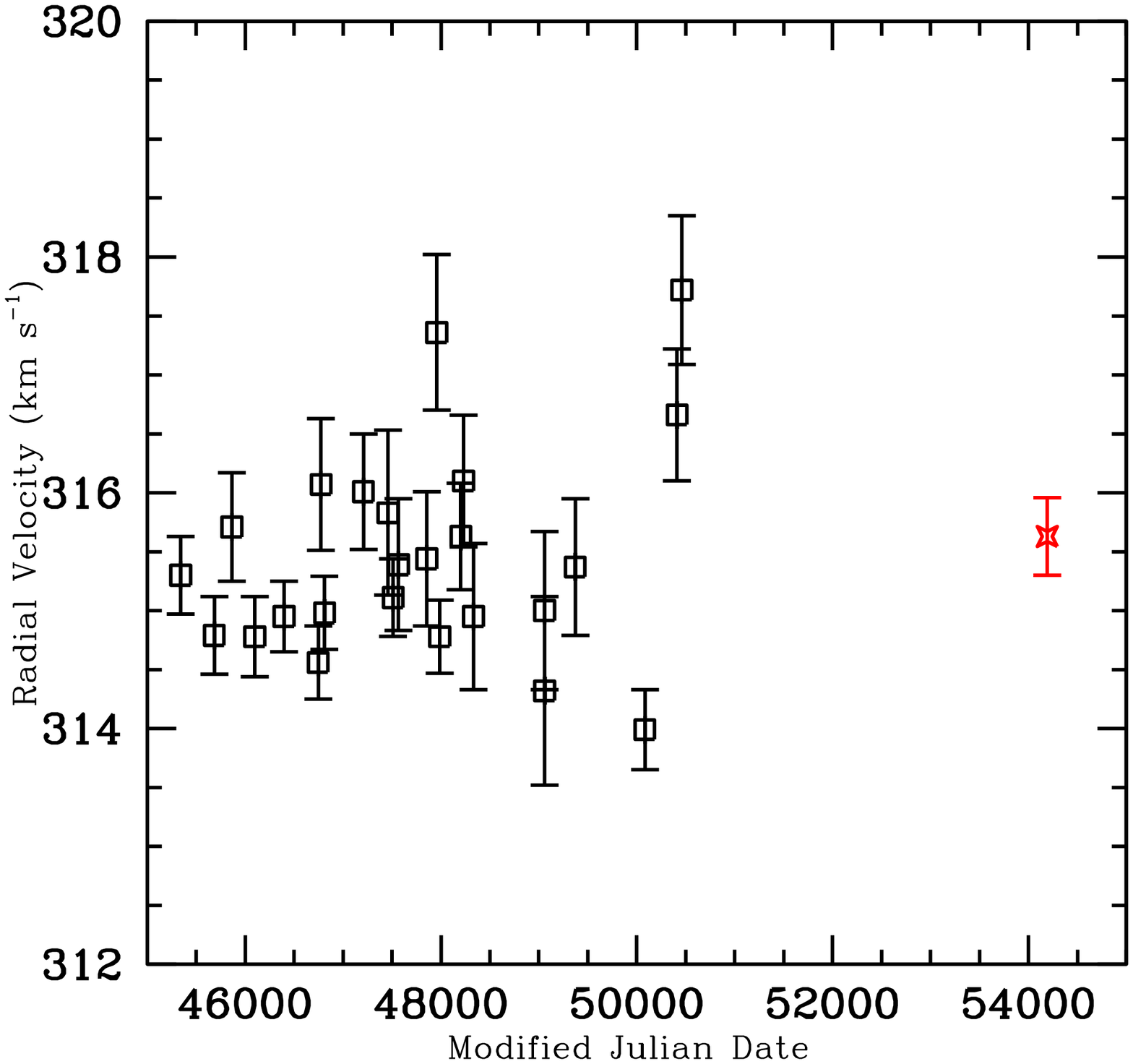}
\caption{Radial velocity of HD 55496 as a function of the Modified
Julian Date (MJD). All data points were taken from Jorissen et al. (2005)
(open squares) except the one at MJD\,=\,54194.052 (red star) which
was taken by us.}
\end{figure}

\subsection{Elemental Abundances}

\par In following Subsections we discuss the abundance pattern of HD
55496 comparing it with previous studies done for some stars in the
halo and globular cluster stars, and also with the other chemically
peculiar stars where heavy-element overabundances have already been
reported in the literature.  Figure 8 shows the abundance pattern of
HD 55496 for all the elements analyzed in this work.

\begin{figure} %Fig8
\includegraphics[width=\columnwidth]{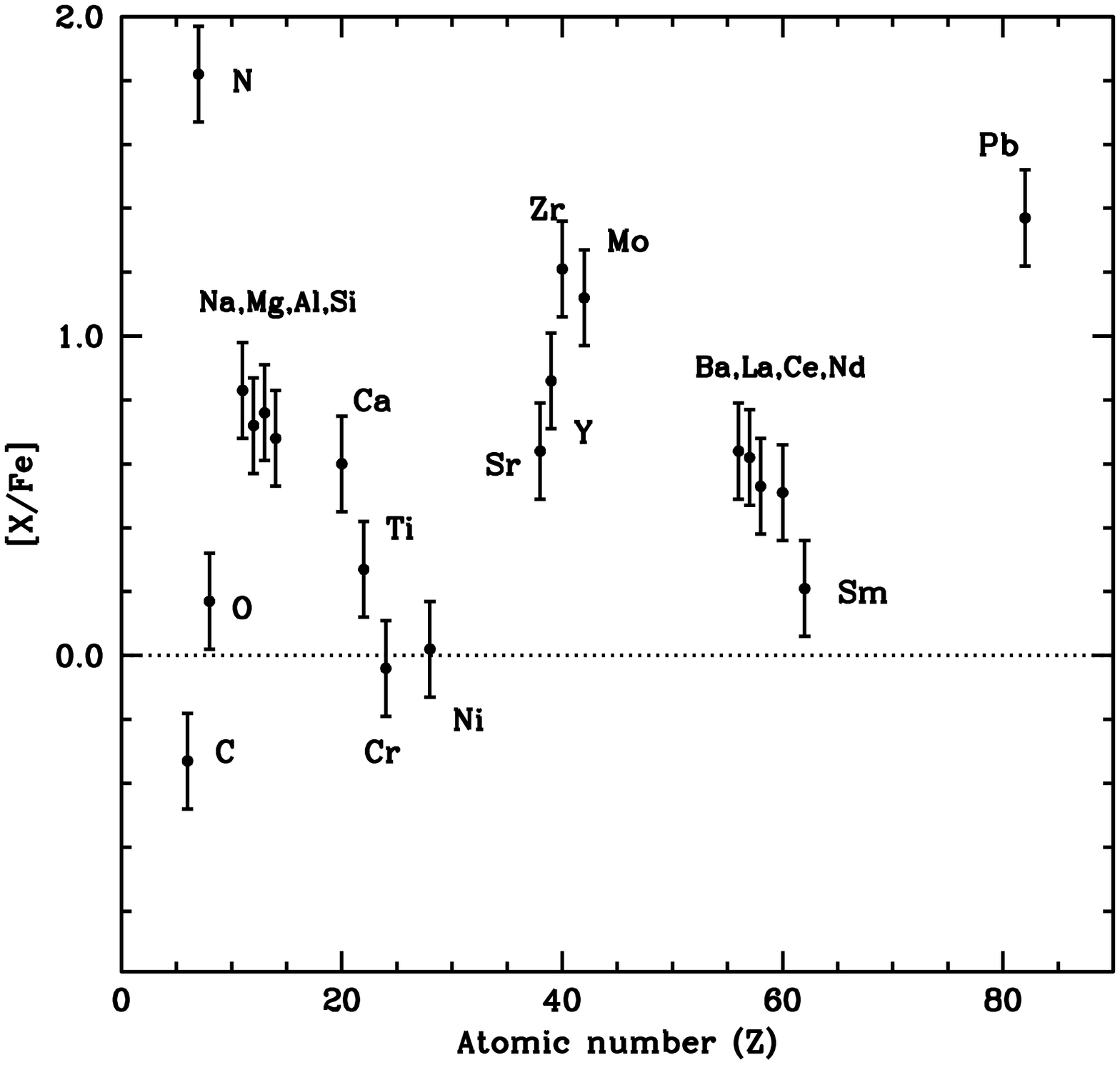}
\caption{Abundance pattern of HD 55496. The mean elemental abundance is shown
versus atomic number}
\end{figure}

\subsubsection{Carbon, Nitrogen, Lithium and
  the $^{12}\mathrm{C}/\,^{13}\mathrm{C}$ isotopic ratio}

\par As shown in Table 4, the carbon abundance is low compared to a
s-process enriched star of similar metallicity such as
CD-62$\degr$1346, with [Fe/H]\,=\,$-$1.59$\pm$0.08 and a [C/Fe]
ratio of $+$0.86 (Pereira et al. 2012). The [N/Fe] ratio is higher
than the field giant stars of similar metallicity (Gratton et
al. 2000).  HD 55496 has a low $^{12}\mathrm{C}/\,^{13}\mathrm{C}$
ratio ($^{12}$C/$^{13}$C\,=\,6.0$\pm$1.0). The high nitrogen abundance
and the low $^{12}\mathrm{C}/\,^{13}\mathrm{C}$ strongly suggest
that significant mixing has occurred in the atmosphere of HD 55496.
In fact, the [N/Fe] ratio {\sl versus} [Fe/H] of dwarf stars shows no
trend in the metallicity range $-2.0 <$ [Fe/H] $<+0.3$, and [N/Fe] is
$\approx 0.0$ (Clegg et al. 1981; Tomkin \& Lambert 1984; Carbon et
al.  1987). As a star becomes a giant, nuclear processed material by
the CN cycle which creates more $^{14}$N and more $^{13}$C relative to
$^{12}$C, due to the deepening of its convective envelope, is brought
from the interior to the outer layers of the star changing the surface
composition.

\par The excess of C$+$N is the difference between the total C$+$N
abundance and the expected primordial value.  In non-evolved stars,
the abundance of nitrogen scales with the iron abundance that is
for metalicities between $+$0.3 and $-$2.0, therefore the [N/Fe]
ratio is basically zero (Wheeler et al. 1989).  For carbon we used the
trend between the [C/H] {\sl versus} [Fe/H] from Masseron et
al. (2006).  In HD 55496, the (C$+$N) abundance sum shows an
enhancement of $\sim$1.1 dex compared to stars with similar
metallicity, which is seen in Figure 9 notwithstanding the low carbon
abundance. The value of (C$+$N) means that the star did not
  experience the triple-alpha process which would result in the carbon
  enhanced (see Table 4).  Therefore, HD 55496 is not at the AGB
phase, a conclusion already obtained in Section 4.1. In addition, the
C/O ratio, which is 0.15, gives another evidence that HD 55496 is not
carbon enriched.

\par Nitrogen can be enhanced through two mixing episodes, cool bottom
processing (CBP) (Wasserburg et al. 1995)\footnote{An extra mixing
  mechanism that takes material from the cool base of the convective
  zone to the deeper layers of the stellar interior where hydrogen
  burning occurs.} or hot bottom burning (HBB) (Sackmann \& Boothroyd
1992)\footnote{Hot Bottom Burning takes place in the stars more
  massive than 4.0 M$_{\odot}$ where the outer convection zone reaches
  temperatures high enough for proton-capture nucleosynthesis 
    occurring at the base of the outer envelope favoring the
  conversion of carbon to nitrogen through the CN-cycle.}  Lithium can
  also be enhanced thanks to these two mixing process, but as seen in
Table 4, lithium is not enhanced.  The absence of a high lithium
abundance rules out CBP, since Sackmann \& Boothroyd (1999) showed
that CBP can account for the high lithium abundances.  Therefore we
should consider HBB as the source of the abundance peculiarities.  In
Section 4.3.2 we will also provide further support for this conclusion
based on the analysis of the abundances of the elements sodium and
aluminum.

\begin{figure} %Fig9
\includegraphics[width=\columnwidth]{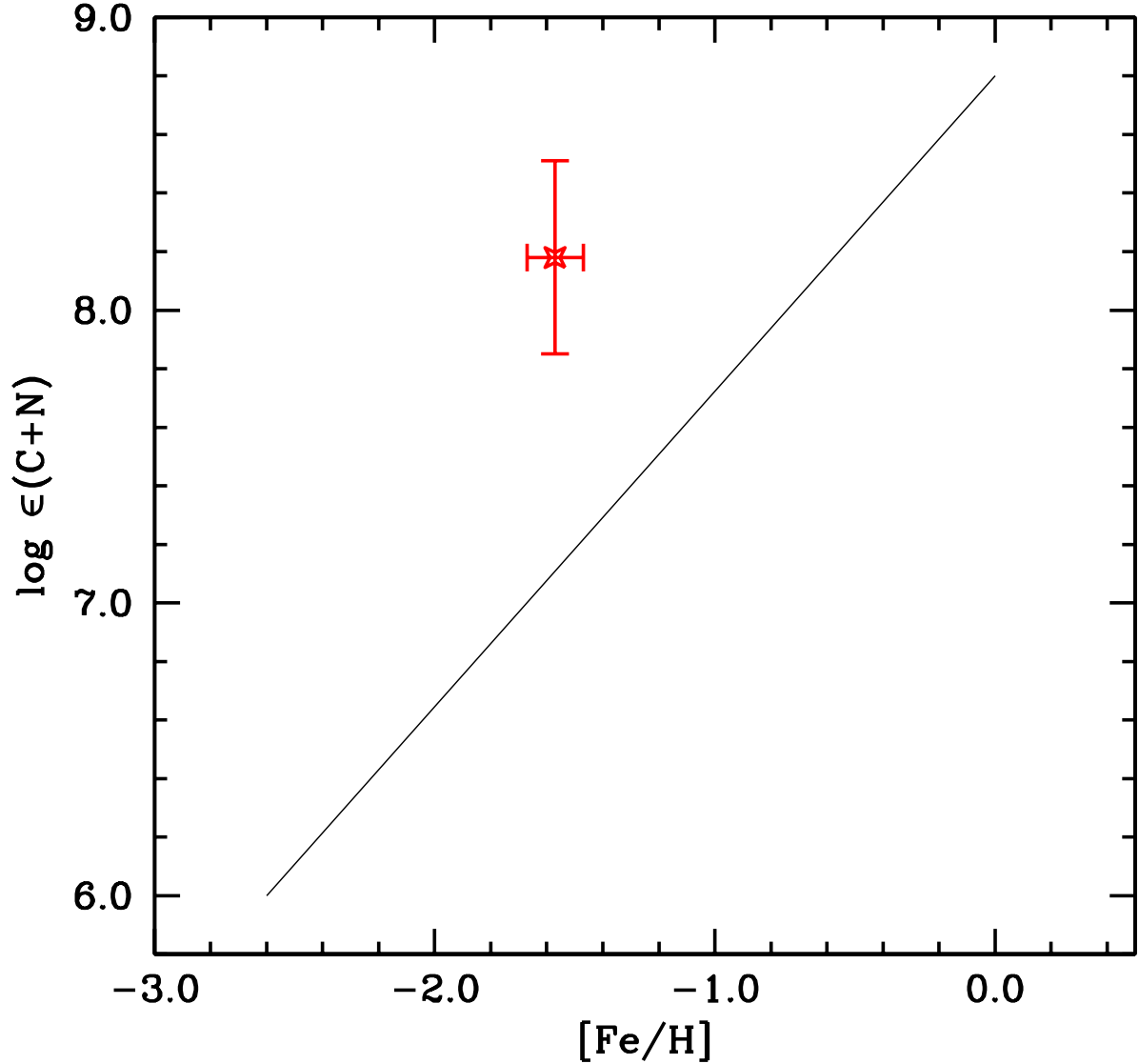}
\caption{Observed C$+$N abundance sum in the notation of log $\epsilon$
(C$+$N). The solid line shows initial CN abundance for a given metallicity. }
\end{figure}

\subsubsection{Sodium, Aluminum, $\alpha$- and iron peak elements}

\par In this Section we comment on the abundances of sodium,
  aluminum elements, alpha-elements and iron-peak elements in comparison with
  field halo stars and globular cluster stars.  Inspecting Table 4,
  the chemical abundances of HD 55496 present an interesting aspect
  concerning the abundance of Na and Al. The abundance of these
  elements is higher compared with field stars of similar metallicity
  (Norris et al. 2001; Carretta et al. 2010).

\par In field halo stars with metallicities down to $-$1.0, the
     [Na/Fe] ratio is not higher than $+$0.5 (Timmes et al. 1995).
     However there are some exceptions, a few stars having [Na/Fe]
     ratios between $+$0.6 and $+$0.7 (Roederer et al. 2014). It is
     therefore possible that they may represent a population
     evaporated from globular clusters.  The abundance of aluminum
     ([Al/Fe]) in extremely metal-poor stars is never higher than
     $+$0.5 (Fullbright 2000; Carretta et al. 2002).  Contrary to
     field stars, both ratios in stars of globular clusters can reach
     ratios higher than $+$1.0 (Shetrone 1996; Carretta et al. 2006,
     just to mention a few references). The [Na/Fe] and [Al/Fe] ratios
     in HD 54496, with values of $+$0.83 and $+$0.76, respectively,
     are similar to ratios usually observed in globular cluster stars
     of the "second generation".  In addition, the [O/Fe] ratio in HD
     55496 is lower than in field stars of similar metallicity.
     Figure 10 shows the position of HD 55496 in the diagram [Na/Fe]
     and [Al/Fe] ratios {\sl versus} the [O/Fe] ratio for some
     globular cluster stars.  Based on the position on these two
     figures, the most likely explanation for the origin of HD 55496,
     is that it is an escaped globular cluster star.

\par As far as the other alpha-elements is concerned, the [Si/Fe],
     [Ca/Fe], and [Ti/Fe] ratios follow the same trend of halo stars
     with similar metallicity. The abundances of chromium and nickel
     follow, as expected, the iron abundance trend. Finally, HD 55496
     shows a [Mg/Fe] ratio of +0.72, which is a value previously found
     in other globular clusters (Gratton et al. 2012; Carretta et
     al. 2013; Gratton et al. 2014; Gratton et al. 2015).

\begin{figure} %Fig10
\includegraphics[width=\columnwidth]{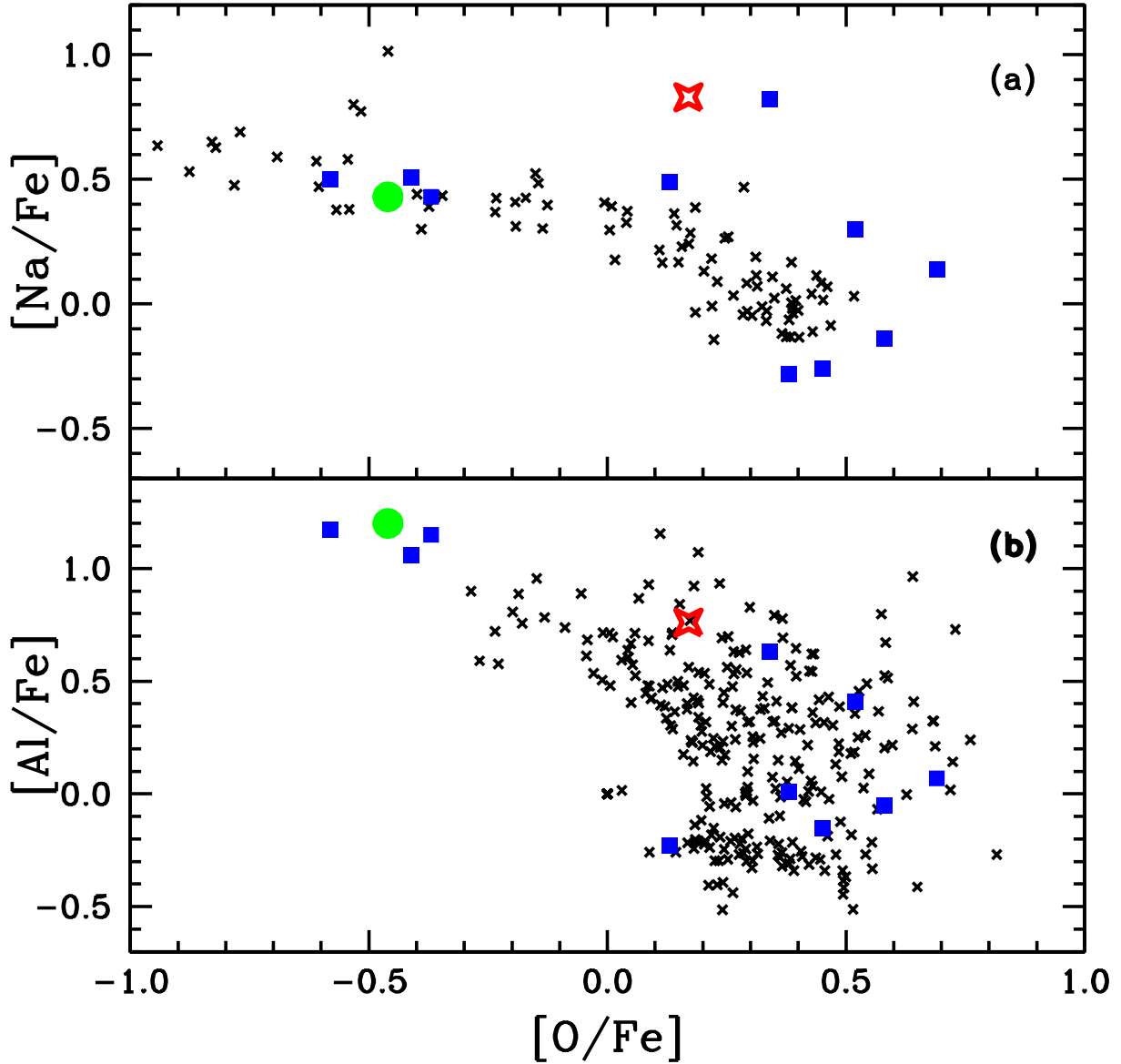}
\caption{Position of HD 55496 (red star) in the [Na/Fe] ratio
{\sl versus} [O/Fe] ratio diagram (a) and in the [Al/Fe] ratio {\sl versus} [O/Fe] ratio
diagram (b), compared to the position TYC 5691-109-1
(green circle, analyzed by Pereira et al. 2017) and also compared to the
position of stars of other globular clusters.
Data were taken from: (i) Carretta et al. (2006) for the [Na/Fe] and [O/Fe] ratios
in NGC 2808 (black crosses); (ii) Fern\'andez-Trincado et al. (2016) for several globular
clusters for the [Al/Fe] and [O/Fe] ratios (black crosses) based on DR13 from APOGEE
and (iii) Smith et al. (2000) for the $\omega$ Cen (blue squares).}
\end{figure}

\subsubsection{The s-process elements}

\par One of the most interesting aspects of the abundance pattern of
HD 55496 for a star to be considered as a ``escaped globular cluster
star'' is the abundance of the s-process elements. From Table 4, it is
evident that the {\sl light} s-process elements (Sr, Y, Zr, and Mo)
show a very high abundance, with a mean of [ls/Fe]\,=\,0.96$\pm$0.26.
The abundances of the {\sl heavy} s-process elements, given by Ba, La,
Ce and Nd, provide a lower [hs/Fe] ratio, which is 0.46$\pm$0.14,
therefore the [hs/ls] ratio is $-$0.50$\pm$0.30.  A similar [hs/ls]
ratio of $-$0.35 was obtained by Karinkuzhi \& Goswami (2015) for HD
55496. In addition, the [s/Fe] ratio ( 's' represents the mean
abundance of the s-process elements Sr, Y, Zr, Mo, Ba, La, Ce and Pb)
is $+$0.80$\pm$0.28. This explains why HD 55496 was previously
classified as a barium star by several authors (MacConnell et
al. 1972; Bond 1974; Catchpole et al. 1977; Lu et al. 1979; Sneden
1983; Luck \& Bond 1991; Karinkuzhi \& Goswami 2015).

\par According to model predictions by Busso et al (2001), for
metallicities between $-$0.5 and $-$2.0 for a 1.5\,$M_{\odot}$ AGB
star and for metallicities between $-$0.5 and $-$1.0 for a
3.0\,$M_{\odot}$ AGB star the [hs/ls] ratio has a maximum value around
$\approx$1.2, which means that heavy s-process elements are the
dominant products of neutron capture in AGB stars at those
metallicities lower than solar. In AGB stars at near solar
metallicities, the light s-process elements are the dominant products
of neutron capture.  Therefore the index [hs/ls] is an important
measure of the neutron capture efficiency and has been widely used in
the AGB nucleosynthesis models.  This anti-correlation between the
[hs/ls] {\sl versus} metallicity is due to the operation of the
$^{13}$C($\alpha,\,n$)$^{16}$O reaction, as this neutron source is
anti-correlated with the metallicity (Wallerstein 1997).

\par Negative [hs/ls] indexes are also predicted by models of Busso et
al. (2001) for stars with $-2.0\le$ [Fe/H] $\le-$1.0 however, none of
the chemically peculiar binary stars where the overabundances of the
elements created by the s-process are due to a mass-transfer during
the AGB phase of the primary star, have negative [hs/ls] index
at the same metallicity of HD 55496.  This is shown in Figure 11 where
the values of the [hs/ls] index for barium stars, CH
stars\footnote{Chemically peculiar binary systems enhanced in carbon
  and s-process elements. CH stars belong to the halo population.}
and CEMP-s stars are displayed.  HD 55496 falls in a different
position compared to the other chemically peculiar binary stars, which
is below the main trend of the anti-correlation between the [hs/ls]
and [Fe/H].  This may raise the possibility that other enrichment
processes, rather than the mass transfer hypothesis, were responsible
for the overabundance in light s-process elements observed in
  HD55496.  Another evidence that can probably rule out the
hypothesis of mass-transfer is due to the inconclusive results of the
behavior of the radial velocity (Section 4.2), not permitting to
confirm or reject a binary nature of HD~55496.  In addition, as seen
in Section 4.1, we may disregard self-enrichment of HD 55496 in
s-process elements because of its low luminosity to be considered an
AGB star.  Therefore, we suggest as the most probable explanation that
the observed overabundance of the light s-process elements was due to
the s-process enriched medium where HD 55496 was born and where AGB
stars have played a dominant role in the chemical evolution.

\par In order to verify this possibility we should search among the
AGB models those that predict negative [hs/ls] index for low
metallicity.  Models of massive AGB stars between 5.0 and
8.0\,M$_{\odot}$ predict negative [hs/ls] ratios considering the
reaction $^{22}$Ne($\alpha$,n)$^{25}$Mg as the most likely neutron
source for the origin of the s-process elements (Karakas \& Lattanzio
2014; Cristallo et al. 2015). In Figure 12 we show the position of HD
55496 in the diagrams [hs/ls] and [Pb/hs] {\sl versus} iron abundance
together with AGB model predictions given by Cristallo et al. (2015)
for AGB stars between 4.0 and 6.0\,M$_{\odot}$ at [Fe/H] $<$
$-$1.0. It is clear that based on the position of HD 55496 in both
diagrams, its abundance pattern can be reproduced by an AGB star with
5.0\,-\,6.0\,M$_{\odot}$.  Therefore, if indeed HD 55496 is a ``second
generation'' escaped globular cluster star, as suggested by the
  overabundances of sodium and aluminum, it is also possible that
  intermediate-mass AGB stars between 3.0 and 6.0\,M$_{\odot}$ were
  responsible for the production of the elements created by s-process
  during the time of formation of HD 55496 in a globular
  cluster.

\begin{figure} %Fig11
\includegraphics[width=\columnwidth]{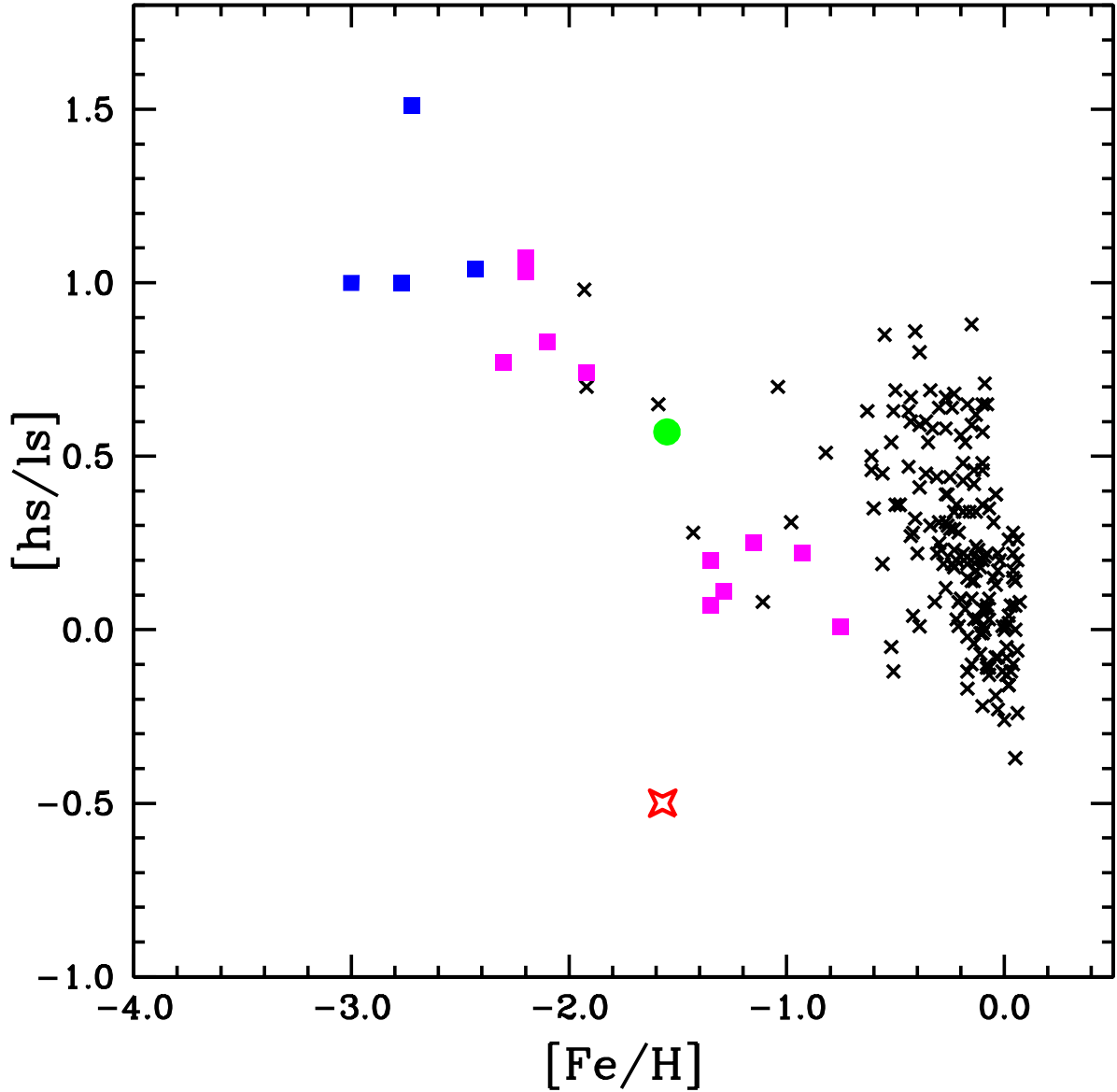}
\caption{[hs/ls] {\sl versus} [Fe/H] for HD 55496
({\it red star}) and several classes of chemically-peculiar binary stars.
Barium giants are represented by {\sl black crosses}, CH stars and yellow
symbiotic stars by {\sl magenta squares} and binary CEMP-s stars by
{\it blue squares}. Data for barium stars, CH stars and yellow symbiotic stars
were taken from de Castro et al. (2016). Data for binary CEMP-s stars
were taken from Preston \& Sneden (2001), Sivarani et al. (2004), Barbuy et
al. (2005), Lucatello et al. (2003) and Thompson et al. (2008).
TYC-5619-109-1 is represented by {\sl green circle}.}
\end{figure}

\begin{figure} %Fig12
\includegraphics[width=\columnwidth]{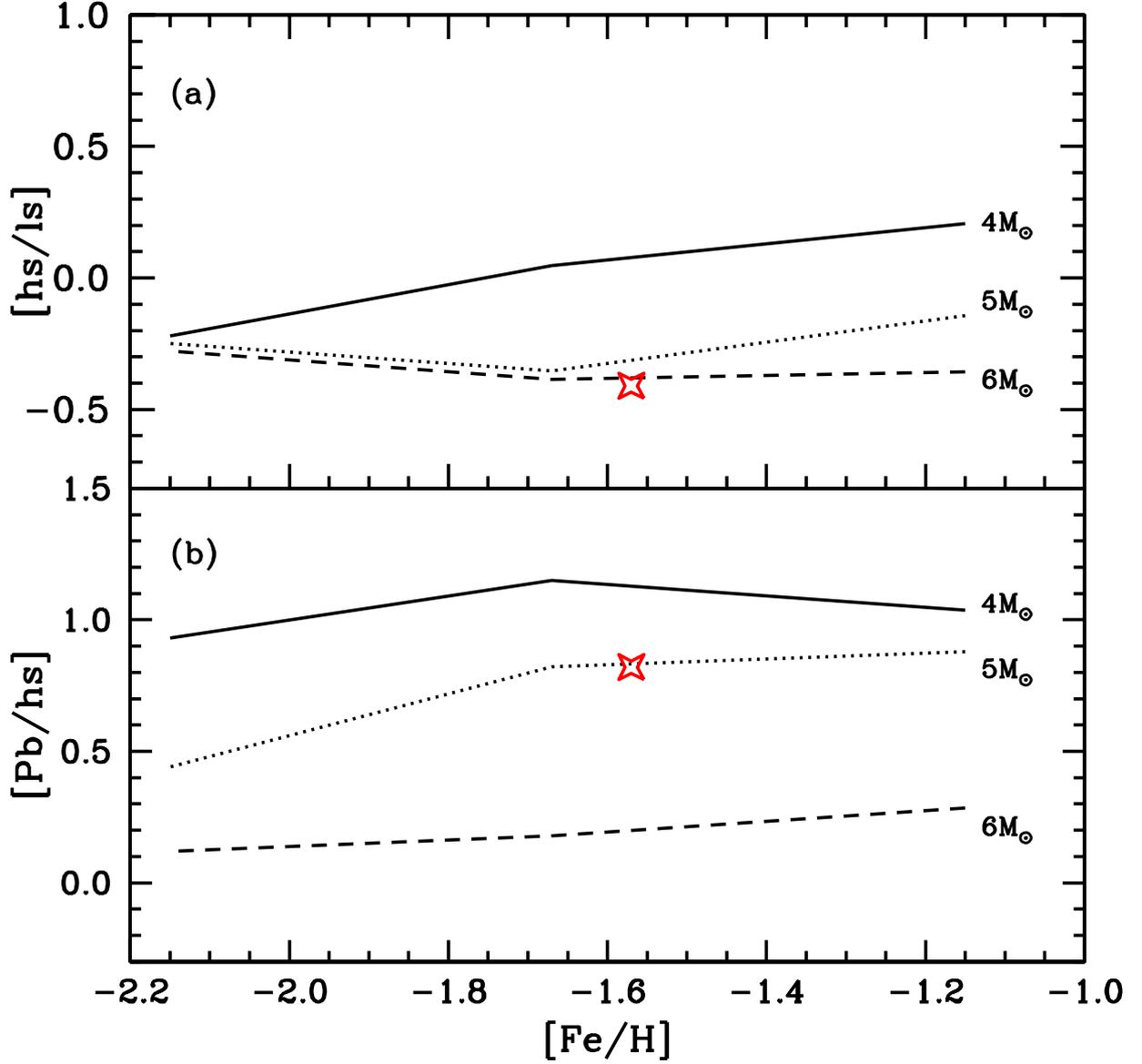}
\caption{[hs/ls] ratio {\sl versus} [Fe/H] (a) and [Pb/hs] ratio {\sl versus} [Fe/H] (b)
for HD 55496 in comparison with AGB model predictions given by Cristallo et al. (2015)
for AGB stars with masses of 4.0\,M$_{\odot}$, 5.0\,M$_{\odot}$ and 6.0\,M$_{\odot}$.}
\end{figure}

\subsection{Kinematics and Dynamical Origin}

\par In order to investigate whether HD 55496 is a bound or unbound
object to the Galaxy and also to investigate the globular cluster
origin of HD 55496 we should first not only compare its velocity in
the Galactocentric Reference Frame system ($V_{\mathrm{GRF}}$) with
the escape Galactic velocity but also to perform a dynamical analysis
of its orbital evolution.

\par Knowning the distance, the proper motion and the radial velocity
of HD 55496 we are able to calculate the space velocity components
($U_{0}$,$V_{0}$,$W_{0}$) using the algorithm by Johnson \& Soderblom
(1987). We also consider the peculiar solar motion of (8.5, 13.4, 6.5)
km\,s$^{-1}$, as derived by Co\c skuno\u glu et al. (2011) and the
proper motion of HD 55496 taken from H\o g et al (2000). A
Galactocentric solar distance of 8.5~kpc and a local standard of rest
(LSR) rotation velocity relative to the GRF of 220~km\,s$^{-1}$ was
adopted in our calculations. The results for $U_{0}$,$V_{0}$,$W_{0}$
are respectively 130.0 km\,s$^{-1}$, $-$55.3km\,s$^{-1}$ and
4.4km\,s$^{-1}$. Considering the uncertainty of 11 pc in the distance
(Section 4.1), we obtain a modulus of the star velocity in the GRF of
$V_{\mathrm{GRF}}$\,=\, 141$\pm$0.5 km\,s$^{-1}$. This velocity is
much lower than the escape Galactic velocity of
$V_{\mathrm{esc}}=532$\,km\,s$^{-1}$, which corresponds to the value
provided by the Galactic gravitational potential of Allen \& Santillan
(1991) at an heliocentric distance of 494 pc.  Therefore HD 55496
appears to be a bound object to the Galaxy, however presents
$V_{0}<0$, which indicate that is a halo intruder object.

\par We performed a numerical simulation of the orbit of HD 55496
using a Bulirsh-Stoer integrator that solves the equations of motion
of the star in the potential of Allen \& Santillan (1991). The star
shows a very elongated trajectory, with an eccentricity of
$e=0.84$. It reaches the priapsis at 0.87~kpc with a velocity of
477~km\,s$^{-1}$, and the apoapsis at 10.45~kpc with a velocity of
40~km\,s$^{-1}$.  In order to check a possible globular cluster origin
of this star, we applied the same procedure as in Pereira et
al. (2017).  The orbit of the star was simulated over 12~Gyr into the
past, together with the orbits of 142 globular clusters from the
catalog of Kharchenko et al. (2013). During the simulation, we tracked
the mutual distances between the star and each cluster, and computed
the encounter probabilities within a given distance $d$. The
simulation was repeated 5000 times considering different orbital
parameters for the star and the clusters within their 1$\sigma$
uncertainties.  Table 8 summarizes the results for values of $d$
corresponding to 1.5 and 5.0 cluster tidal radii. The low encounter
probabilities indicate that the origin of HD~55496 is unlikely to be
related to any globular cluster. The few cases displaying the largest
probabilities are also associated to very large mutual encounter
velocities, indicating that even in such cases the star probably did
not detach from the cluster by tidal forces.  From the dynamical point
of view, HD~55496 has similarities with TYC~5619-109-1 (Pereira et
al. 2017), and may represent another example of a possible leftover
from a tidally disrupted dwarf galaxy.

\begin{table}%T8
\caption{Probabilities for HD 55496 to have a close encounter with a globular
cluster within a given distance in terms of the cluster 
tidal radius, $r_\mathrm{c}$. Only the four clusters with the largest probabilities 
are reported, together with the corresponding average encounter velocities
$v_\mathrm{enc}$. Cluster iron abundances are from Kharchenko et al. (2013)}\label{tabprob}
\begin{tabular}{lcccc}\hline\hline
Name   & $p(<1.5r_\mathrm{c})$ & $p(<5r_\mathrm{c})$ & $v_\mathrm{enc}$ & [Fe/H] \\
       &    (\%)    &    (\%)  &  km\,s$^{-1}$  &     \\\hline
NGC 5139  &  6.0   &   16.2  &  $370\pm 120$  &  -1.445 \\
NGC 6402  &  4.0   &   14.8  &  $502\pm 157$  &  -1.145 \\
NGC 6749  &  4.6   &   16.3  &  $528\pm 204$  &  -1.600 \\
Terzan 4  &  3.4   &   15.3  &  $589\pm 99$   &  -1.045 \\
\hline
\end{tabular}
\end{table}

\section{Conclusions}

\par Based on high-resolution optical spectroscopic data, we presented
a new chemical analysis of HD 55496, a star previously known as barium
star. We determined abundances of light elements, Na, Al,
alpha-elements, iron-peak elements, and s-process elements.  We showed
that HD 55496 is a metal-poor star characterized by an enhancement of
nitrogen, sodium, aluminum and s-process elements including lead.  The
oxygen abundance is low when compared to field metal poor stars of
similar metallicity.

\par Our results show that HD 55496 present a Na--O anti-correlation
usually observed among globular cluster stars. The star is also
aluminum rich.  HD 55496 presents a ratio [Al/Mg]\,=\,$+$0.04 which
could be classified as a value between to a primordial and
intermediate globular cluster populations according to Carretta et
al. (2012). But since the [Al/Fe] ratio is $+$0.76, which is a
``typical'' value for an intermediate population, HD 55496 owed,
probably, its abundance pattern due to a different polluters (Carretta
et al. 2012).

\par As was classified as a barium star, HD 55496 is s-process
enriched but its heavy-element abundance pattern strongly differs from
those of the low-metallicity and chemically-peculiar binary stars, in
sense that the abundance of Sr, Y and Zr is higher than the abundance
of Ba, La, Ce and Nd.  In fact, the value of the [hs/ls] ratio is
$-$0.50, never observed for a chemically-peculiar binary star at the
metallicity of HD 55496.  This suggests that the main source of the
neutrons is not the $^{13}$C($\alpha,\,n$)$^{16}$O reaction but the
$^{22}$Ne($\alpha$,n)$^{25}$Mg reaction. If this is the case, the
origin of the heavy-element abundance pattern was provided by stars
with masses higher than 4.0\,M$_{\odot}$, or by the intermediate mass
stars which have polluted HD 55496, just like these stars were
responsible for the abundances seen among the ``second generation of
the globular cluster stars''.  In the earlier paper (Pereira et
al. 2017) we showed that TYC 5619-109-1 presented a chemistry that may
have been formed in a globular cluster.  In addition, it was also
showed that TYC 5619-109-1 is carbon poor and s-process
enriched. However, HD 55496 differs from TYC 5619-109-1 in sense that
the HD 55496 seems to owe its heavy-element abundance pattern due a
pollution by massive stars while for TYC 5619-109-1 the contamination
may have happened by a mass transfer or by a pollution of a gas
already strongly enriched in s-process elements.  Both HD 55496 and
TYC 5619-109-1 are up to now the only two known s-process enriched
globular cluster escaped stars.  Finally, the kinematical and
dynamical analysis shows that HD 55496 may have originated from a
globular cluster however having a retrograde motion, we can not rule
out the possibility that may have originated from a tidally disrupted
dwarf galaxy. The low encounter probabilities with a globular clusters
leaves open the origin of HD 55496.

\section{Acknowledgements}

\par We thank Verne Smith, Simone Daflon and Jo\~ao Victor S. Silva
for fruitful discussions. NAD acknowledges Russian Foundation for
Basic Research (RFBR) according to the research projects 18-02-00554
and 18-52-06004. Finally we thank the referee for the suggestions and
comments that improved the quality of the paper.

{}

\appendix

\begin{table*}%T1
\caption{Observed Fe\,{\sc i} and Fe\,{\sc ii} lines}
\begin{tabular}{|l|c|c|c|c|}\hline
Element & $\lambda$(\AA) & $\chi$(eV) & $\log gf$ & W$_{\lambda}$(m\AA) \\\hline
Fe\,{\sc i}\, &   5133.69   &   4.18  &    0.20   &    89\\
 &   5150.84   &   0.99  &   $-$3.00   &   114\\
 &   5151.91   &   1.01  &   $-$3.32   &   106\\
 &   5159.06   &   4.28  &   $-$0.65   &    30\\
 &   5162.27   &   4.18  &      0.07   &    82\\
 &   5194.94   &   1.56  &   $-$2.09   &   123\\
 &   5198.71   &   2.22  &   $-$2.14   &    87\\
 &   5232.94   &   2.94  &   $-$0.08   &   139\\
 &   5242.49   &   3.63  &   $-$0.97   &    58\\
 &   5250.21   &   0.12  &   $-$4.92   &    85\\
 &   5281.79   &   3.04  &   $-$0.83   &   101\\
 &   5302.31   &   3.28  &   $-$0.74   &    90\\
 &   5307.36   &   1.61  &   $-$2.97   &    84\\
 &   5322.04   &   2.28  &   $-$2.84   &    43\\
 &   5339.93   &   3.27  &   $-$0.68   &    95\\
 &   5341.02   &   1.61  &   $-$1.95   &   135\\
 &   5364.87   &   4.45  &      0.23   &    71\\
 &   5367.47   &   4.42  &      0.43   &    75\\
 &   5369.96   &   4.37  &      0.54   &    82\\
 &   5373.71   &   4.47  &   $-$0.71   &    22\\
 &   5393.17   &   3.24  &   $-$0.72   &    91\\
 &   5400.50   &   4.37  &   $-$0.10   &    59\\
 &   5410.91   &   4.47  &      0.40   &    71\\
 &   5445.04   &   4.39  &      0.04   &    62\\
 &   5487.75   &   4.32  &   $-$0.65   &    43\\
 &   5506.78   &   0.99  &   $-$2.80   &   130\\
 &   5522.45   &   4.21  &   $-$1.40   &    14\\
 &   5554.90   &   4.55  &   $-$0.38   &    41\\
 &   5560.21   &   4.43  &   $-$1.04   &    15\\
 &   5563.60   &   4.19  &   $-$0.84   &    46\\
 &   5567.39   &   2.61  &   $-$2.56   &    38\\
 &   5569.62   &   3.42  &   $-$0.49   &    95\\
 &   5572.84   &   3.40  &   $-$0.28   &   104\\
 &   5576.09   &   3.43  &   $-$0.85   &    76\\
 &   5624.02   &   4.39  &   $-$1.33   &    12\\
 &   5686.53   &   4.55  &   $-$0.45   &    26\\
 &   5691.50   &   4.30  &   $-$1.37   &    10\\
 &   5717.83   &   4.28  &   $-$0.97   &    25\\
 &   5731.76   &   4.26  &   $-$1.15   &    22\\
 &   5762.99   &   4.21  &   $-$0.41   &    58\\
 &   5791.02   &   3.21  &   $-$2.27   &    36\\
 &   5883.82   &   3.96  &   $-$1.21   &    34\\
 &   5916.25   &   2.45  &   $-$2.99   &    37\\
 &   6024.06   &   4.55  &   $-$0.06   &    53\\
 &   6027.05   &   4.08  &   $-$1.09   &    29\\
 &   6056.01   &   4.73  &   $-$0.40   &    25\\
 &   6065.48   &   2.61  &   $-$1.53   &   106\\
 &   6079.01   &   4.65  &   $-$0.97   &    25\\
 &   6136.61   &   2.45  &   $-$1.40   &   115\\
 &   6137.69   &   2.59  &   $-$1.40   &   107\\
 &   6151.62   &   2.18  &   $-$3.29   &    31\\
 &   6157.73   &   4.08  &   $-$1.11   &    36\\
 &   6170.51   &   4.79  &   $-$0.38   &    28\\
 &   6173.34   &   2.22  &   $-$2.88   &    59\\
 &   6191.56   &   2.43  &   $-$1.42   &   114\\
 &   6200.31   &   2.60  &   $-$2.44   &    51\\
 &   6213.43   &   2.22  &   $-$2.48   &    74\\
 &   6230.72   &   2.56  &   $-$1.28   &   121\\
 &   6252.56   &   2.40  &   $-$1.72   &   107\\
 &   6265.13   &   2.18  &   $-$2.55   &    85\\
 &   6322.69   &   2.59  &   $-$2.43   &    59\\
 &   6393.60   &   2.43  &   $-$1.43   &   120\\
 &   6411.65   &   3.65  &   $-$0.66   &    82\\
 &   6419.95   &   4.73  &   $-$0.09   &    41\\
 &   6421.35   &   2.28  &   $-$2.01   &   100\\
 &   6430.85   &   2.18  &   $-$2.01   &   108\\
 &   6592.91   &   2.72  &   $-$1.47   &    93\\
 &   6593.87   &   2.44  &   $-$2.42   &    68\\\hline
\end{tabular}
\end{table*}

\begin{table*} 
\noindent{Table 1, continued}
\begin{tabular}{|l|c|c|c|c|}\hline
Element & $\lambda$(\AA) & $\chi$(eV) & $\log gf$ & W$_{\lambda}$(m\AA) \\\hline
 &   6609.11   &   2.56  &   $-$2.69   &    49\\
 &   6750.15   &   2.42  &   $-$2.62   &    69\\\hline
Fe\,{\sc ii}\,  &  4993.35  &  2.81 &  $-$3.67 &   26\\
 &  5132.66  &  2.81 &  $-$4.00 &   12\\
 &  5197.56  &  3.23 &  $-$2.25 &   66\\
 &  5284.10  &  2.89 &  $-$3.01 &   43\\
 &  5325.56  &  3.22 &  $-$3.17 &   24\\
 &  5414.05  &  3.22 &  $-$3.62 &   15\\
 &  5425.25  &  3.20 &  $-$3.21 &   26\\
 &  5534.83  &  3.25 &  $-$2.77 &   40\\
 &  6084.10  &  3.20 &  $-$3.80 &   11\\
 &  6432.68  &  2.89 &  $-$3.58 &   29\\\hline
\end{tabular}
\end{table*}

\addtocounter{table}{1}
\begin{table*}%T3
\caption{Other lines studied}
\begin{tabular}{|l|c|c|c|c|c|}\hline
$\lambda$(\AA) & Element & $\chi$(eV) & $\log gf$ & Ref & W$_{\lambda}$(m\AA) \\\hline
4664.81  & \ph{1}Na\,{\sc i} & 2.10 & $-$1.55 & T94 &  28 \\
4668.60  & \ph{1}Na\,{\sc i} & 2.10 & $-$1.30 & T94 &  29 \\
4982.81  & \ph{1}Na\,{\sc i} & 2.10 & $-$0.95 & T94 &  54 \\
5682.65  & \ph{1}Na\,{\sc i} & 2.10 & $-$0.70 & T94 &  77 \\ 
5688.22  & \ph{1}Na\,{\sc i} & 2.10 & $-$0.37 & T94 &  97 \\ 
6154.22  & \ph{1}Na\,{\sc i} & 2.10 & $-$1.51 & T94 &  23 \\ 
6160.75  & \ph{1}Na\,{\sc i} & 2.10 & $-$1.21 & T94 &  40 \\  
8183.26  & \ph{1}Na\,{\sc i} & 2.10 & $+$0.22 & T94 & 168 \\\hline   

4702.99  & \ph{1}Mg\,{\sc i} & 4.35 & $-$0.52 & A2004     & 146 \\   
4730.04  & \ph{1}Mg\,{\sc i} & 4.34 & $-$2.39 & R03       &  35 \\   
5528.40  & \ph{1}Mg\,{\sc i} & 4.35 & $-$0.49 & A2004     & 157 \\   
5711.10  & \ph{1}Mg\,{\sc i} & 4.34 & $-$1.68 & R99       &  78 \\   
8736.04  & \ph{1}Mg\,{\sc i} & 5.94 & $-$0.34 & WSM\ph{1} &  82 \\\hline

6696.03  & \ph{1}Al\,{\sc i} &  3.14 &  $-$1.48 & MR94 &  18\\
6698.67  & \ph{1}Al\,{\sc i} &  3.14 &  $-$1.63 & R03  &  17\\
7835.32  & \ph{1}Al\,{\sc i} &  4.02 &  $-$0.58 & R03  &  28\\
7836.13  & \ph{1}Al\,{\sc i} &  4.02 &  $-$0.40 & R03  &  25\\
8772.88  & \ph{1}Al\,{\sc i} &  4.02 &  $-$0.25 & R03  &  41\\
8773.91  & \ph{1}Al\,{\sc i} &  4.02 &  $-$0.07 & R03  &  59\\\hline

5793.08  & \ph{1}Si\,{\sc i} & 4.93 & $-$2.06 & R03 &    46 \\
6125.03  & \ph{1}Si\,{\sc i} & 5.61 & $-$1.54 & E93 &    15 \\
6131.58  & \ph{1}Si\,{\sc i} & 5.62 & $-$1.69 & C2003 &  10 \\
6145.08  & \ph{1}Si\,{\sc i} & 5.62 & $-$1.43 & E93 &    43 \\
7800.00  & \ph{1}Si\,{\sc i} & 6.18 & $-$0.72 & E93 &    27 \\
8742.45  & \ph{1}Si\,{\sc i} & 5.87 & $-$0.51 & E93 &    65 \\\hline
5581.80  & \ph{1}Ca\,{\sc i} & 2.52 & $-$0.67 & C2003 &  76 \\
5601.29  & \ph{1}Ca\,{\sc i} & 2.52 & $-$0.52 & C2003 &  77 \\
5857.46  & \ph{1}Ca\,{\sc i} & 2.93 & $+$0.11 & C2003 &  99 \\
6102.73  & \ph{1}Ca\,{\sc i} & 1.88 & $-$0.79 & D2002 & 113 \\
6122.23  & \ph{1}Ca\,{\sc i} & 1.89 & $-$0.32 & D2002 & 142 \\
6166.44  & \ph{1}Ca\,{\sc i} & 2.52 & $-$1.14 & R03   &  48 \\
6169.04  & \ph{1}Ca\,{\sc i} & 2.52 & $-$0.80 & R03   &  74 \\
6169.56  & \ph{1}Ca\,{\sc i} & 2.53 & $-$0.48 & DS91  &  88 \\
6439.08  & \ph{1}Ca\,{\sc i} & 2.52 & $+$0.47 & D2002 & 137 \\
6449.82  & \ph{1}Ca\,{\sc i} & 2.52 & $-$0.50 & C2003 &  85 \\
6455.60  & \ph{1}Ca\,{\sc i} & 2.51 & $-$1.29 & R03   &  40 \\
6464.68  & \ph{1}Ca\,{\sc i} & 2.52 & $-$2.41 & C2003 &  13 \\
6471.66  & \ph{1}Ca\,{\sc i} & 2.51 & $-$0.69 & S86   &  78 \\
6493.79  & \ph{1}Ca\,{\sc i} & 2.52 & $-$0.11 & DS91  & 113 \\
6499.65  & \ph{1}Ca\,{\sc i} & 2.52 & $-$0.81 & C2003 &  72 \\
6717.69  & \ph{1}Ca\,{\sc i} & 2.71 & $-$0.52 & C2003 &  84 \\\hline
\end{tabular}
\end{table*}

\begin{table*}
\noindent{Table 3, continued}
\bigskip
\begin{tabular}{|l|c|c|c|c|c|}\hline
$\lambda$(\AA) & Element & $\chi$(eV) & $\log gf$ & Ref & W$_{\lambda}$(m\AA)\\\hline
4758.12  & \ph{1}Ti\,{\sc i} &  2.25 &  $+$0.43  & MFK  &   29\\
4759.28  & \ph{1}Ti\,{\sc i} &  2.25 &  $+$0.51  & MFK  &   32\\
4981.74  & \ph{1}Ti\,{\sc i} &  0.85 &  $+$0.50  &D2002 &  115\\
4997.10  & \ph{1}Ti\,{\sc i} &  0.00 &  $-$2.12  & MFK  &   48\\
5009.66  & \ph{1}Ti\,{\sc i} &  0.02 &  $-$2.26  & MFK  &   35\\
5016.17  & \ph{1}Ti\,{\sc i} &  0.85 &  $-$0.57  & MFK  &   75\\
5022.87  & \ph{1}Ti\,{\sc i} &  0.83 &  $-$0.43  & MFK  &   79\\
5039.96  & \ph{1}Ti\,{\sc i} &  0.02 &  $-$1.13  & MFK  &   94\\
5043.59  & \ph{1}Ti\,{\sc i} &  0.84 &  $-$1.73  & MFK  &   16\\
5087.06  & \ph{1}Ti\,{\sc i} &  1.43 &  $-$0.84  & E93  &   22\\
5145.47  & \ph{1}Ti\,{\sc i} &  1.46 &  $-$0.57  & MFK  &   29\\
5147.48  & \ph{1}Ti\,{\sc i} &  0.00 &  $-$2.01  & MFK  &   52\\
5152.19  & \ph{1}Ti\,{\sc i} &  0.02 &  $-$2.02  & MFK  &   51\\
5210.39  & \ph{1}Ti\,{\sc i} &  0.05 &  $-$0.88  & MFK  &  108\\
5219.71  & \ph{1}Ti\,{\sc i} &  0.02 &  $-$2.29  & MFK  &   34\\
5295.78  & \ph{1}Ti\,{\sc i} &  1.05 &  $-$1.63  & MFK  &   14\\
5490.16  & \ph{1}Ti\,{\sc i} &  1.46 &  $-$0.94  & MFK  &   18\\
5866.46  & \ph{1}Ti\,{\sc i} &  1.07 &  $-$0.87  & E93  &   51\\
5922.12  & \ph{1}Ti\,{\sc i} &  1.05 &  $-$1.47  & MFK  &   22\\
5978.55  & \ph{1}Ti\,{\sc i} &  1.87 &  $-$0.50  & MFK  &   20\\
6126.22  & \ph{1}Ti\,{\sc i} &  1.05 &  $-$1.37  & R03  &   22\\
6258.11  & \ph{1}Ti\,{\sc i} &  1.44 &  $-$0.36  & MFK  &   47\\
6261.10  & \ph{1}Ti\,{\sc i} &  1.43 &  $-$0.48  & MFK  &   41\\
6554.24  & \ph{1}Ti\,{\sc i} &  1.44 &  $-$1.22  & MFK  &   18\\\hline
4789.34  & \ph{1}Cr\,{\sc i} &  2.54 &  $-$0.33  & S07  &   44\\
4829.37  & \ph{1}Cr\,{\sc i} &  2.54 &  $-$0.51  & S07  &   21\\
4870.80  & \ph{1}Cr\,{\sc i} &  3.08 &  $-$0.01  & S07  &   29\\
5296.70  & \ph{1}Cr\,{\sc i} &  0.98 &  $-$1.36  & S07  &   76\\
5298.28  & \ph{1}Cr\,{\sc i} &  0.98 &  $-$1.14  & S07  &   87\\
5345.81  & \ph{1}Cr\,{\sc i} &  1.00 &  $-$0.95  & S07  &   97\\
5409.79  & \ph{1}Cr\,{\sc i} &  1.03 &  $-$0.67  & S07  &  109\\
5783.87  & \ph{1}Cr\,{\sc i} &  3.32 &  $-$0.29  & S07  &   10\\
5787.93  & \ph{1}Cr\,{\sc i} &  3.32 &  $-$0.08  & S07  &   11\\
6330.10  & \ph{1}Cr\,{\sc i} &  0.94 &  $-$2.92  & S07  &   14\\\hline
4519.98  & \ph{1}Ni\,{\sc i} &  1.68 &  $-$3.08  & W2014 &  22\\
4604.99  & \ph{1}Ni\,{\sc i} &  3.48 &  $-$0.24  & W2014 &  51\\
4715.76  & \ph{1}Ni\,{\sc i} &  3.54 &  $-$0.33  & W2014 &  42\\
4756.52  & \ph{1}Ni\,{\sc i} &  3.48 &  $-$0.27  & W2014 &  46\\
4866.27  & \ph{1}Ni\,{\sc i} &  3.54 &  $-$0.22  & W2014 &  48\\
4953.21  & \ph{1}Ni\,{\sc i} &  3.74 &  $-$0.58  & W2014 &  26\\
4976.33  & \ph{1}Ni\,{\sc i} &  1.68 &  $-$3.00  & W2014 &  27\\
5003.75  & \ph{1}Ni\,{\sc i} &  1.68 &  $-$3.07  & W2014 &  24\\
5010.94  & \ph{1}Ni\,{\sc i} &  3.63 &  $-$0.98  & W2014 &  21\\
5017.58  & \ph{1}Ni\,{\sc i} &  3.54 &  $-$0.03  & W2014 &  56\\
5035.36  & \ph{1}Ni\,{\sc i} &  3.64 &  $+$0.29  & W2014 &  64\\
5084.10  & \ph{1}Ni\,{\sc i} &  3.68 &  $-$0.18  & W2014 &  51\\
5102.97  & \ph{1}Ni\,{\sc i} &  1.68 &  $-$2.87  & W2014 &  32\\
5115.40  & \ph{1}Ni\,{\sc i} &  3.83 &  $-$0.28  & W2014 &  38\\
5424.65  & \ph{1}Ni\,{\sc i} &  1.95 &  $-$2.74  & W2014 &  26\\
5435.86  & \ph{1}Ni\,{\sc i} &  1.99 &  $-$2.58  & W2014 &  35\\\hline
\end{tabular}
\end{table*}

\begin{table*}
\noindent{Table 3, continued}
\bigskip
\begin{tabular}{|l|c|c|c|c|c|}\hline
$\lambda$(\AA) & Element & $\chi$(eV) & $\log gf$ & Ref & W$_{\lambda}$(m\AA)\\\hline
5578.73 & \ph{1}Ni\,{\sc i}  &  1.68 &  $-$2.67  & W2014 &  43\\
5587.87 & \ph{1}Ni\,{\sc i}  &  1.94 &  $-$2.39  & W2014 &  42\\
5592.26 & \ph{1}Ni\,{\sc i}  &  1.95 &  $-$2.51  & W2014 &  44\\
5709.56 & \ph{1}Ni\,{\sc i}  &  1.68 &  $-$2.30  & W2014 &  73\\
5748.36 & \ph{1}Ni\,{\sc i}  &  1.68 &  $-$3.24  & W2014 &  19\\
5846.99 & \ph{1}Ni\,{\sc i}  &  1.68 &  $-$3.46  & W2014 &  16\\
5892.87 & \ph{1}Ni\,{\sc i}  &  1.99 &  $-$2.22  & W2014 &  58\\
6007.31 & \ph{1}Ni\,{\sc i}  &  1.68 &  $-$3.40  & W2014 &  13\\
6108.12 & \ph{1}Ni\,{\sc i}  &  1.68 &  $-$2.49  & W2014 &  54\\
6128.98 & \ph{1}Ni\,{\sc i}  &  1.68 &  $-$3.32  & W2014 &  18\\
6176.82 & \ph{1}Ni\,{\sc i}  &  4.09 &  $-$0.26  & W2014 &  23\\
6177.25 & \ph{1}Ni\,{\sc i}  &  1.83 &  $-$3.51  & W2014 &  10\\
6327.60 & \ph{1}Ni\,{\sc i}  &  1.68 &  $-$3.15  & W2014 &  29\\
6482.80 & \ph{1}Ni\,{\sc i}  &  1.94 &  $-$2.63  & W2014 &  29\\
6532.88 & \ph{1}Ni\,{\sc i}  &  1.94 &  $-$3.39  & W2014 &  11\\
6586.33 & \ph{1}Ni\,{\sc i}  &  1.95 &  $-$2.81  & W2014 &  24\\
6643.64 & \ph{1}Ni\,{\sc i}  &  1.68 &  $-$2.03  & W2014 &  86\\
6767.77 & \ph{1}Ni\,{\sc i}  &  1.83 &  $-$2.17  & W2014 &  72\\
6772.32 & \ph{1}Ni\,{\sc i}  &  3.66 &  $-$0.97  & W2014 &  17\\
7788.93 & \ph{1}Ni\,{\sc i}  &  1.95 &  $-$2.18  & W2014 &  75\\\hline

4607.33 & Sr\,{\sc i} & 0.00 & $+$0.28 & SN96 &   70 \\\hline

4883.68 & Y\,{\sc ii} & 1.08 & $+$0.07 & SN96 &  105 \\
5087.43 & Y\,{\sc ii} & 1.08 & $-$0.17 & SN96 &   94 \\
5123.21 & Y\,{\sc ii} & 0.99 & $-$0.93 & SN96 &   71 \\
5200.41 & Y\,{\sc ii} & 0.99 & $-$0.57 & SN96 &   82 \\
5205.72 & Y\,{\sc ii} & 1.03 & $-$0.34 & SN96 &  101 \\
5289.81 & Y\,{\sc ii} & 1.03 & $-$1.85 & VWR  &   24 \\
5402.78 & Y\,{\sc ii} & 1.84 & $-$0.44 & R03  &   38 \\\hline

4772.30 & \ph{1}Zr\,{\sc i} & 0.62 & $-$0.06 & A04 &  24 \\
4815.05 & \ph{1}Zr\,{\sc i} & 0.65 & $-$0.38 & A04 &  21 \\
5385.13 & \ph{1}Zr\,{\sc i} & 0.52 & $-$0.64 & A04 &  11 \\
6127.46 & \ph{1}Zr\,{\sc i} & 0.15 & $-$1.06 & S96 &  15 \\
6134.57 & \ph{1}Zr\,{\sc i} & 0.00 & $-$1.28 & S96 &  12 \\
6143.18 & \ph{1}Zr\,{\sc i} & 0.07 & $-$1.10 & S96 &  19 \\\hline

4050.32 & \ph{1}Zr\,{\sc ii} & 0.71 &  $-$1.06 & L2006  & 79\\     
4442.99 & \ph{1}Zr\,{\sc ii} & 1.49 &  $-$0.42 & L2006  & 69\\    
5112.27 & \ph{1}Zr\,{\sc ii} & 1.66 &  $-$0.85 & L2006  & 47\\   
5350.35 & \ph{1}Zr\,{\sc ii} & 1.77 &  $-$1.16 & L2006  & 31\\\hline

5570.44 & \ph{1}Mo\,{\sc i} &  1.34 &  $-$0.34 & Y2017 & 10\\\hline

4086.71 & \ph{1}La\,{\sc ii} & 0.00 & $-$0.16 & SN96  &  79\\
5880.63 & \ph{1}La\,{\sc ii} & 0.24 & $-$1.83 & VWR   &  15\\
6774.33 & \ph{1}La\,{\sc ii} & 0.12 & $-$1.71 & VWR   &  18\\\hline

\end{tabular}
\end{table*}

\begin{table*}
\noindent{Table 3, continued}
\bigskip
\begin{tabular}{|l|c|c|c|c|c|}\hline
$\lambda$(\AA) & Element & $\chi$(eV) & $\log gf$ & Ref & W$_{\lambda}$(m\AA)\\\hline
4418.79 & \ph{1}Ce\,{\sc ii} & 0.86 & $+$0.27 & L09  & 41\\
4483.90 & \ph{1}Ce\,{\sc ii} & 0.86 & $+$0.10 & L09  & 29\\
4486.91 & \ph{1}Ce\,{\sc ii} & 0.29 & $-$0.18 & L09  & 39\\
4539.74 & \ph{1}Ce\,{\sc ii} & 0.33 & $-$0.08 & L09  & 49\\
4562.37 & \ph{1}Ce\,{\sc ii} & 0.48 & $+$0.21 & L09  & 59\\
4628.16 & \ph{1}Ce\,{\sc ii} & 0.52 & $+$0.14 & L09  & 52\\
5187.45 & \ph{1}Ce\,{\sc ii} & 1.21 & $+$0.17 & L09  & 17\\
5274.24 & \ph{1}Ce\,{\sc ii} & 1.04 & $+$0.13 & L09  & 26\\
5330.58 & \ph{1}Ce\,{\sc ii} & 0.87 & $-$0.40 & L09  & 11\\
5393.39 & \ph{1}Ce\,{\sc ii} & 1.10 & $-$0.06 & L09  & 14\\\hline

4462.98  & \ph{1}Nd\,{\sc ii} & 0.56 &  $+$0.04 & DH  & 53\\
4706.54  & \ph{1}Nd\,{\sc ii} & 0.00 &  $-$0.71 & DH  & 39\\
4709.72  & \ph{1}Nd\,{\sc ii} & 0.18 &  $-$0.97 & DH  & 26\\
4715.59  & \ph{1}Nd\,{\sc ii} & 0.20 &  $-$0.90 & DH  & 17\\
4797.15  & \ph{1}Nd\,{\sc ii} & 0.56 &  $-$0.69 & DH  & 13\\
4825.48  & \ph{1}Nd\,{\sc ii} & 0.18 &  $-$0.42 & DH  & 44\\
4859.03  & \ph{1}Nd\,{\sc ii} & 0.32 &  $-$0.44 & DH  & 38\\
4902.04  & \ph{1}Nd\,{\sc ii} & 0.06 &  $-$1.34 & DH  & 15\\
4914.38  & \ph{1}Nd\,{\sc ii} & 0.38 &  $-$0.70 & DH  & 20\\
4959.12  & \ph{1}Nd\,{\sc ii} & 0.06 &  $-$0.80 & DH  & 31\\
5092.80  & \ph{1}Nd\,{\sc ii} & 0.38 &  $-$0.61 & DH  & 26\\
5130.59  & \ph{1}Nd\,{\sc ii} & 1.30 &  $+$0.45 & DH  & 25\\
5165.13  & \ph{1}Nd\,{\sc ii} & 0.68 &  $-$0.74 & DH  & 15\\
5212.36  & \ph{1}Nd\,{\sc ii} & 0.20 &  $-$0.96 & DH  & 21\\
5234.19  & \ph{1}Nd\,{\sc ii} & 0.55 &  $-$0.51 & DH  & 22\\
5249.58  & \ph{1}Nd\,{\sc ii} & 0.98 &  $+$0.20 & DH  & 30\\
5255.51  & \ph{1}Nd\,{\sc ii} & 0.20 &  $-$0.67 & DH  & 35\\
5293.16  & \ph{1}Nd\,{\sc ii} & 0.82 &  $+$0.10 & DH  & 28\\
5311.46  & \ph{1}Nd\,{\sc ii} & 0.98 &  $-$0.42 & DH  & 12\\
5319.81  & \ph{1}Nd\,{\sc ii} & 0.55 &  $-$0.14 & DH  & 37\\ 
5485.70  & \ph{1}Nd\,{\sc ii} & 1.26 &  $-$0.12 & DH  & 13\\\hline

4318.94  & \ph{1}Sm\,{\sc ii} &  0.28 &  $-$0.25 & L06  & 27\\
4424.32  & \ph{1}Sm\,{\sc ii} &  0.48 &     0.14 & L06  & 36\\
4467.34  & \ph{1}Sm\,{\sc ii} &  0.66 &     0.15 & L06  & 21\\
4642.23  & \ph{1}Sm\,{\sc ii} &  0.38 &  $-$0.46 & L06  & 16\\
4676.90  & \ph{1}Sm\,{\sc ii} &  0.04 &  $-$0.87 & L06  & 14\\
4704.40  & \ph{1}Sm\,{\sc ii} &  0.00 &  $-$0.86 & L06  & 17\\\hline

\end{tabular}
\par {\bf References :} A2004 : Aoki et al. (2004); A04: Antipova et al. (2004);
\par C2003: Chen et al. (2003); D2002: Depagne et al. (2002);
\par DH: Den Hartog et al. (2003); DS91: Drake \& Smith (1991);
\par E93: Edvardsson et al. (2003); L06: Lawler et al. (2006);
\par L2006: Ljung et al. (2006); L09: Lawler et al. (2009);
\par MFK : Martin et al. (2002); R03: Reddy et al. (2003);
\par R99: Reddy et al. (1999); S86: Smith et al. (1986);
\par S96: Smith et al. (1996); SN96: SN96: Sneden et al. (1996);
\par S07: Sobeck et al. (2007); T94: Takeda \& Takada-Hidai (1994);
\par WSM: Wiese, Smith \& Miles (1969); W2014: Wood et al. (2014);
\par Y2017: Yong et al. (2017);

\end{table*}

\bsp
\label{lastpage}
\end{document}